\documentclass[sn-mathphys,iicol]{sn-jnl}% Default with double column layout

\usepackage{graphicx}
\usepackage{physics}
\usepackage{booktabs}
\usepackage{adjustbox}
\usepackage{amsmath,amsfonts,amssymb}
\usepackage[labelfont=bf]{subcaption}
\jyear{2021}%

\def\vec#1{\ensuremath{\mathchoice
           {\mbox{\boldmath$\displaystyle\mathbf{#1}$}}
           {\mbox{\boldmath$\textstyle\mathbf{#1}$}}
           {\mbox{\boldmath$\scriptstyle\mathbf{#1}$}}
           {\mbox{\boldmath$\scriptscriptstyle\mathbf{#1}$}}}}%

\newcommand{\F}{\vec F} %Deformation Gradient
\newcommand{\E}{\vec E} %Left Cauchy Green Stress
\def\P{\vec S} %Second Piola Kirchoff stress

\def\V{V_\text{m}} %transmembrane potential
\def\w{\vec w} % Gating variables and ion concentrations
\def\c{\vec c} % ion concentrations
\def\Cm{C_\text{m}} % membrane capacitance
\def\DT{\boldsymbol\sigma} %Conductivity Tensor
\def\Iion{I_{\rm{ion}}} %ionic current
\def\Iext{I_{\rm{ext}}} % external stimulus
\newcommand{\OmegaEP}{\Omega_\mathrm{EP}}
\newcommand{\OmegaM}{\Omega_\mathrm{M}}

\DeclareMathSymbol{\Gamma}{\mathalpha}{operators}{0}
\DeclareMathSymbol{\Delta}{\mathalpha}{operators}{1}
\DeclareMathSymbol{\Theta}{\mathalpha}{operators}{2}
\DeclareMathSymbol{\Lambda}{\mathalpha}{operators}{3}
\DeclareMathSymbol{\Xi}{\mathalpha}{operators}{4}
\DeclareMathSymbol{\Pi}{\mathalpha}{operators}{5}
\DeclareMathSymbol{\Sigma}{\mathalpha}{operators}{6}
\DeclareMathSymbol{\Upsilon}{\mathalpha}{operators}{7}
\DeclareMathSymbol{\Phi}{\mathalpha}{operators}{8}
\DeclareMathSymbol{\Psi}{\mathalpha}{operators}{9}
\DeclareMathSymbol{\Omega}{\mathalpha}{operators}{10}

\theoremstyle{thmstyleone}%
\theoremstyle{thmstyletwo}%

\theoremstyle{thmstylethree}%

\begin{document}

\title[ ]{The Impact of Standard Ablation Strategies for Atrial Fibrillation on Cardiovascular Performance in a Four-chamber Heart Model}


\author*[1]{\fnm{Tobias} \sur{Gerach}}\email{publications@ibt.kit.edu}

\author[1]{\fnm{Steffen} \sur{Schuler}}

\author[1]{\fnm{Andreas} \sur{Wachter}}

\author[1]{\fnm{Axel} \sur{Loewe}}

\affil[1]{\orgdiv{Institute of Biomedical Engineering}, \orgname{Karlsruhe Institute of Technology (KIT)}, \orgaddress{\city{Karlsruhe}, \country{Germany}}}


\abstract{
\textbf{Purpose:} Atrial fibrillation is one of the most frequent cardiac arrhythmias in the industrialized world and ablation therapy is the method of choice for many patients.
However, ablation scars alter the electrophysiological activation and the mechanical behavior of the affected atria.
Different ablation strategies with the aim to terminate atrial fibrillation and prevent its recurrence exist but their impact on the hemodynamic performance of the heart has not been investigated thoroughly. \\
\textbf{Methods:} In this work, we present a simulation study analyzing five commonly used ablation scar patterns and their combinations in the left atrium regarding their impact on the pumping function of the heart using an electromechanical whole-heart model.
We analyzed how the altered atrial activation and increased stiffness due to the ablation scar affect atrial as well as ventricular contraction and relaxation.\\
\textbf{Results:} We found that systolic and diastolic function of the left atrium is impaired by ablation scars and that the reduction of atrial stroke volume of up to 11.43\% depends linearly on the amount of inactivated tissue.
Consequently, the end-diastolic volume of the left ventricle, and thus stroke volume, was reduced by up to 1.4\% and 1.8\%, respectively.
During ventricular systole, left atrial pressure was increased by up to 20\% due to changes in the atrial activation sequence and the stiffening of scar tissue.\\
\textbf{Conclusion:} This study provides biomechanical evidence that atrial ablation has acute effects not only on atrial contraction but also on ventricular pumping function. 
Our results have the potential to help tailoring ablation strategies towards minimal global hemodynamic impairment.
}




\maketitle

\section{Introduction}
\label{introduction}

Although atrial fibrillation (AF) can often be managed pharmacologically, catheter ablation is now considered as a first line treatment for paroxysmal AF patients~\citep{Hindricks-2020-ID14630}. 
Unfortunately, the long-term success rates of catheter ablation of AF are not satisfactory yet.
One reason could be long term adaptations of the atria to compensate for ablation-induced impairment of atrial function.
Thomas et al.~\cite{Thomas2000} analyzed the effect of atrial radiofrequency ablation (RFA) on the atrial mechanical function. 
They concluded that multiple linear RFA lesions may impair atrial contractility and that the reduced atrial function is partly due to loss of viable myocardial tissue but may also be caused by the altered atrial activation.
Different ablation strategies with the aim to prevent AF and its recurrence exist but their respective impacts on the ventricular mechanics and hemodynamics have not been studied systematically. 
In this context, cardiac modeling opens up the possibility to get a better understanding of how RFA affects the mechanical behavior and the pumping function of the heart by applying different treatment strategies to the same virtual patient under controlled conditions, which is not feasible in a clinical context.

While in the past most attention in the field of cardiac modeling had been paid to the simulation of the ventricles, and still is an active field of research \citep{Gurev-2015-ID15963,Pfaller-2018-ID12429, Regazzoni-2021-ID16493,Shavik-2021-ID16187, Augustin-2021-ID16580}, modeling of the atria has moved into focus in the last years. 
However, most publications deal with atrial electrophysiology and there are only a few addressing atrial mechanics \citep{Doessel2012}. 
Jernigan et al.~\cite{Jernigan2007} published a study on the mechanical properties of porcine left atrium, which were assessed with uniaxial tests. 
In 2011, DiMartino et al.~\cite{DiMartino2011a} presented a computational model of the porcine left atrium, which allowed to analyze the wall stress due to the mitral valve movement, obtained from raw multi-detector computed tomography data. 
The mechanical properties were modeled based on biaxial experiments with porcine atrial tissue. 
In the same year, DiMartino et al.~\cite{DiMartino2011b} analyzed how ventricular tachypacing affects the spatial and temporal stress distribution of the left atrial wall. 
Bellini et al.~\cite{Bellini2012} provided a comprehensive characterization of the passive biomechanics of the left human atria based on a Fung-type elastic strain energy potential. 
These publications all deal with the passive mechanical behavior of the atria.
Models for the active contraction of the atria have only been published recently \citep{Land-2018-ID11737, Augustin-2020-ID14141} and the simulation of whole heart models is becoming more feasible~\citep{Gerach-2021}.
A recent study by Hormann et al.~\cite{Hormann-2017-ID12950} modeled atrial systole using standard ablation strategies in a non-pathological and a dilated atrium to study the impact of ablation lesions on the mechanical performance of the atria.
Their model was able to detect differences in left atrial contractility and ejection fraction for multiple activation sequences resulting from RFA and atrial fibrillation-induced atrial remodelling represented by a reduction in global conduction velocity.
Since they were only interested in atrial systole, the ventricles were only represented by a passive stiffness and they only included a Windkessel to model the hemodynamics of the atria, thus neglecting a change in pre- and afterload over multiple heart beats.
%%%%%%%%
Another study that dealt with atrial mechanics was published by Phung et al.~\cite{Phung-2017}.
Unlike Hormann et al.~\cite{Hormann-2017-ID12950}, they connected the left atrium to a circulation model for more realistic hemodynamic conditions and validated their baseline model against pre-treatment data by means of left atrial wall motion.
Phung et al. found that ablating the posterior wall had a smaller impact on atrial function compared to other ablation patterns such as pulmonary vein isolation or wide area circumferential ablation.
They explained this with the little movement of this region in the pre-treatment model and thus small contribution to atrial function.
A limitation of the study by Phung et al. is the missing representation of electrical propagation in the model.
%%%%%%%%

We hypothesize that ablation scars in the atria also have an effect on the ventricles. 
During ventricular contraction, the atrioventricular plane is pulled towards the apex, the atria are stretched and their volume is increased.
This mechanism supports the filling of the atria with blood from the venae cavae and the pulmonary veins. 
Eventually, this blood volume is available for ventricular filling during ventricular relaxation. 
An increased stiffness of the atria due to ablation scars may impede atrioventricular plane displacement (AVPD) and consequently also atrial filling. 
Accordingly, two effects may affect the ventricular filling negatively: On the one hand, a reduced filling capacity of the atria during the ventricular systole and therefore less blood volume available for the ventricular filling.  
On the other hand, a reduced active contribution of the atrium due to an altered atrial activation induced by the ablation scars. 
Here, the effect certainly depends on the respective ablation pattern and the amount of ablated tissue.
Alhogbani et al.~\cite{Alhogbani13} assessed the contribution of the left atrium to the filling of the left ventricle using cardiac magnetic resonance imaging (MRI) for 120 normal subjects. 
They reported that the contribution of the atria to ventricular filling increases with age and is in the range of $10\% - 40\%$.
Considering that about $70\%$ of the patients with AF are between 65 and 85 years old \citep{Fuster2011}, the implications of RFA on the atrial contraction should not be disregarded.
To overcome the limitations of the studies by Hormann et al.~\cite{Hormann-2017-ID12950,Phung-2017}, we utilize our previously published four-chamber heart model including a closed-loop circulatory system, electrophysiological wave propagation and myocardial contraction \citep{Gerach-2021} to extend a previous study from our group \citep{busch13} by analyzing the impact of five commonly used ablation lesions on cardiovascular performance.
Furthermore, we analyze how the stiffening of scar tissue due to an accumulation of collagen in the myocardium affects the deformation of the heart and in particular AVPD.
%%%%%%%%
The sensitivity of the model towards changes in the stiffness of the scars is evaluated for one of the scar patterns.
Finally, we introduce the global effects of AF related fibrosis by reducing the diffusion coefficient.
%%%%%%%%

\section{Methods}
\label{methods}

\subsection{Whole heart finite element model}

The cardiac anatomy was manually segmented from magnetic resonance imaging (MRI) data ($0.72 \times 0.72 \times 1.8\,\text{mm}^3$) of a 33 year old male volunteer. The volunteer provided informed consent and the study was approved by the IRB of Heidelberg University Hospital \citep{fritz13a}.
The MRI data were acquired using a 1.5\,T MR tomography system and consist of a static whole heart image stack taken during diastasis as well as time-resolved images in several long and short axis slices.
Based on the segmentation, we first labeled the atria ($\Omega_\mathrm{A} = \Omega_\mathrm{LA} \cup \Omega_\mathrm{RA}$) and the ventricles ($\Omega_\mathrm{V} = \Omega_\mathrm{LV} \cup \Omega_\mathrm{RV}$).
As shown in \cite{Gerach-2021}, we extended the geometry with a representation of the mitral valve, the tricuspid valve, the aortic valve and the pulmonary valve ($\Omega_\mathrm{Valves}$).
Additionally, we closed the endo- and epicardial surfaces of the atria and added truncated pulmonary veins, vena cavae as well as the ascending aorta and pulmonary artery ($\Omega_\mathrm{Vessels}$).
Compared to the study in \cite{Gerach-2021}, no adipose tissue in the atrioventricular and interventricular grooves was included due to numerical issues caused by locally insufficient element quality.
Furthermore, we added a concentric layer of tissue around the entire heart ($\Omega_\mathrm{Peri}$) which phenomenologically represents the influence of the pericardium \citep{fritz13a}.

Two tetrahedral meshes were created using Gmsh \citep{Geuzaine-2009-ID12650}: (1) the mechanical reference domain $\OmegaM =  \Omega_\mathrm{V} \cup \Omega_\mathrm{A} \cup \Omega_\mathrm{Valves} \cup \Omega_\mathrm{Vessels} \cup \Omega_\mathrm{Peri}$ with 128,976 elements and (2) the electrophysiological reference domain $\OmegaEP = \Omega_\mathrm{V} \cup \Omega_\mathrm{A}$ as a subset of $\OmegaM$ with 50,058,295 elements.
In case of the mechanical mesh, we made sure that the model featured a minimum of two elements throughout the wall of the heart.
We used rule-based methods to assign the myofiber orientation on $\OmegaEP$ as a local basis $\vec Q = (\vec f_0, \vec s_0, \vec n_0)$ for the atria \citep{wachter15,andreas_wachter_2021_4738369} and the ventricles \citep{bayer12,ldrb-fibers}.
The fiber angle in the ventricles was chosen as $60^\circ$ and $-60^\circ$ on the endocardial and epicardial surface, respectively. 
The sheet angle was set to $-65^\circ$ on the endocardium and $25^\circ$ on the epicardium \citep{bayer12}.
For the mechanical mesh, we interpolated the myocyte orientation from the vertices of $\OmegaEP$ to the quadrature points of $\OmegaM$.

\subsection{Electromechanical PDE model}

The electromechanical model used in this study to simulate a four-chamber model of a human heart comprises three main components.
First, we have the electrical propagation model driven by cellular electrophysiology.
Second, we describe the solid mechanics model that is used to simulate the active and passive forces in the myocardium.
Last, the mechanical model is extended by boundary conditions, which originate from physiologically important mechanisms such as the circulatory system or the tissue surrounding the heart.
A complete description of the fully coupled electromechanical model was recently given in Gerach et al~\cite{Gerach-2021}.
Since the focus of this study is on mechanical features and not on the subtleties of mechanoelectric feedback, we only implement excitation-contraction coupling mechanisms and forego the effects of deformation on electrical propagation.

\subsubsection{Electrical propagation model}

The electrical propagation in the myocardium is modeled by the monodomain equation on the reference domain $\OmegaEP \times (0,T]$:
\begin{equation}
    \beta \Cm	\partial_t \V + \beta \Iion(\V, \w, \c) = \grad(\DT \grad{\V}) + \beta \Iext \,,
    \label{eq:PDEep}
\end{equation}
with the transmembrane voltage $\V$, the cellular \linebreak surface-to-volume ratio $\beta$, the membrane capacitance $\Cm$, the total ionic current $\Iion$, and an externally applied stimulus current $\Iext$.
The anisotropic conductivity tensor
\begin{equation}
	\DT(\vec x) = 
	\sigma_\text{f} \,\vec f_0 \otimes \vec f_0 +
	\sigma_\text{s} \,\vec s_0 \otimes \vec s_0 +
	\sigma_\text{n} \,\vec n_0 \otimes \vec n_0 \,,
\end{equation}
depends on the fiber $\vec f_0$, sheet $\vec s_0$, and sheet-normal $\vec n_0$ orientation of the myocardium in the reference configuration with generally higher conductivity values in the fiber direction $\sigma_\text{f}\geq \sigma_\text{s}\geq \sigma_\text{n}\geq 0$.
We set the conductivity in the atria such that the total atrial activation takes approximately 100\,ms \citep{nielsen15}.
For the ventricles, we tuned the conductivities according to Mendonca et al.~\cite{mendonca13} to match conduction velocities of 0.6\,m/s, 0.4\,m/s, 0.2\,m/s in fiber, sheet, and sheet-normal directions \citep{Augustin-2016-ID11752}.

The transmembrane voltage $\V$ is coupled to a system of ordinary differential equations, where $\w$~describes a state vector of different gating mechanisms and $\c$ describes the vector of intracellular ion concentrations.
To account for the inter-chamber differences in electrical activity, we use the model of Courtemanche et al.~\cite{courtemanche98} in the atria and the model of O'Hara et al.~\cite{ohara11} in the ventricles.
The O'Hara et al. model was further modified with the changes proposed by \cite{Passini-2016-ID14034} and \cite{Dutta-2017-ID12202} to ensure robust conduction.
For the implementation and parameterization of these models, we refer to the original publications.

The external stimulus $\Iext(\vec x, t)$ is applied at the junction of the right atrial appendage and the superior vena cava~\citep{loewe16e} to initiate the depolarization wave in the atria.
In the ventricles, we use consistent biventricular coordinates \citep{Schuler-2021-ID15766} to define seven positions (Table \ref{tab:fascicles}) that are known to be common sites of earliest activation \citep{durrer70, Cardone-Noott-2016-ID13338}.
To account for the conduction delay introduced by the atrioventricular node, the ventricles are stimulated 160\,ms after the atria.
Additionally, a thin, fast-conducting endocardial layer is defined in the ventricles to mimic the His-Purkinje system.
We choose the conduction velocity in this layer to be twice as high as in the ventricular bulk tissue.
\begin{table*}[ht]
\caption{Sites of earliest activation in terms of the coordinate system Cobiveco~\citep{Schuler-2021-ID15766}: $a$~apicobasal; $m$~transmural; $r$~rotational; $v$~transventricular. $\delta_\mathrm{m}$ and $\delta_\mathrm{rad}$ are the transmural and radial extent of the activation site, respectively.}
\label{tab:fascicles}
\centering
%\adjustbox{max width=0.48\textwidth}{
\begin{tabular}{lccl}
\toprule
\textbf{Root point} & \multicolumn{2}{c}{\textbf{Extent}} \\
$\vec x_\mathrm{root} = \{ a, m, r, v \}$ & $\delta_\mathrm{m}$ & $\delta_\mathrm{rad}$ & Fascicle description \\
\midrule
$\{0.62, 1, 0.075, 0\}$  & $0.05$ & $3$\,mm & LV mid-posterior superior \\
$\{0.55, 1, 0.175, 0\}$  & $0.05$ & $3$\,mm & LV mid-posterior inferior\\
$\{0.85, 1, 0.57, 0\}$   & $0.05$ & $3$\,mm & LV basal anterior paraseptal\\
$\{0.40, 1, 0.825, 0\}$  & $0.05$ & $3$\,mm & LV mid septal\\
$\{0.45, 1, 0.825, 1\}$  & $0.05$ & $3$\,mm & RV mid septal\\
$\{0.80, 1, 0.44, 1\}$   & $0.05$ & $3$\,mm & RV free wall anterior\\
$\{0.85, 1, 0.25, 1\}$   & $0.05$ & $3$\,mm & RV free wall posterior\\
\bottomrule
\end{tabular}
\end{table*}
Finally, the electrophysiological model is complemented by homogeneous Neumann boundary conditions on the entire surface $\Gamma_\mathrm{EP}$.
All parameters of the electrical propagation model are given in Table~\ref{tab:EPParameter}.

Equation~\eqref{eq:PDEep} is solved numerically using linear tetrahedral finite elements in space and a first order operator splitting in time with explicit integration for the reaction term and implicit integration for the diffusion term ($\Delta t_\mathrm{EP} = 10$\,\textmu s).

\subsubsection{Active and passive mechanics}

The biomechanical behavior of the heart is described by the balance of linear momentum equation in its total Lagrangian formulation
\begin{equation}
    \rho_0 \partial^2_t\vec u -\div(\F\,\P (\vec x, \F )) = \vec 0 \qin \OmegaM \times (0,T] \,, \label{eq:linearMomentum}
\end{equation}
where the deformation tensor $\F = \vec I + \grad_{\vec X}{\vec u}$ describes the deformation from the reference configuration $\OmegaM = \Omega_0(\vec X)$ into the current configuration $\Omega_t(\vec x)$, $\vec u$ is the displacement vector, $\rho_0$ the mass density, and $\P$ is the second Piola-Kirchhoff stress.
To account for the active stress generated during cardiac contraction, we additively decompose $\P$ into passive and active components, such that
\begin{equation}
    \P = \P_\text{pas} + \P_\text{act}\,.
\end{equation}
Assuming that the myocardium is an hyperelastic and nearly incompressible material ($J:=\det(\F) \approx 1$), we can describe the passive stress response with
\begin{equation}
    \P_\text{pas} = 2\pdv{\Psi(\vec C)}{\vec C} = \pdv{\Psi(\vec E)}{\vec E} \,,
\end{equation}
adopting the orthotropic strain energy function $\Psi (\vec E)$ from \cite{Usyk-2000-ID14569}:
\begin{alignat}{1}
    &\Psi_\mathrm{myo}(\E) = 
        \frac{\mu}{2}(\exp(\alpha Q) -1) + \frac{\kappa}{2} (\log{J})^2 \,, \nonumber \\
    &Q = b_\text{ff} E_\text{ff}^2 + b_\text{ss} E_\text{ss}^2 + b_\text{nn} E_\text{nn}^2
         + b_\text{fs} (E_\text{fs}^2 + E_\text{sf}^2) \nonumber \\
    &\qquad  + b_\text{fn} (E_\text{fn}^2 + E_\text{nf}^2) + b_\text{ns} (E_\text{ns}^2 + E_\text{sn}^2) \,,\label{eq:usyk}
\end{alignat}
where $\mu$ and $\kappa \gg \mu$ are the shear and bulk modulus, respectively. 
$E_\text{ij} = \vec E \vec i_0 \cdot \vec j_0$ for $i,j \in \{f,s,n\}$ are the entries of the Green-Lagrange strain tensor ${\vec E = \frac{1}{2}(\vec C - \vec I)}$ with the right Cauchy-Green deformation tensor $\vec C = \F^\mathrm{T}\F$.
We keep the anisotropic scaling factors $b_\mathrm{ij}$ fixed using the values $b_\mathrm{ff} = 1$, $b_\mathrm{ss} = 0.4$, $b_\mathrm{nn} = 0.3$, $b_\mathrm{fs} = 0.7$, $b_\mathrm{fn} = 0.6$, $b_\mathrm{ns} = 0.2$ and only adjust $\mu$ and $\alpha$.
$\Omega_\mathrm{Valves}$, $\Omega_\mathrm{Vessels}$, and $\Omega_\mathrm{Peri}$ are purely passive tissue without a preferred direction of orientation.
Therefore, we use a Neo-Hookean model for these materials:
\begin{equation}
    \Psi_\mathrm{NH}(\vec C) = \frac{\mu}{2} (\tr\vec{\hat{C}} - 3) + \frac{\kappa}{2} (J - 1)^2 \,, \label{eq:NH}
\end{equation}
with $\vec{\hat{C}} = J^{-2/3} \vec C$.
All constitutive law parameters are given in Table~\ref{tab:PassiveMech}.
%%% Mech parameters
\begin{table}[ht]
\caption{Choice of parameters for the constitutive law models given by equations \eqref{eq:usyk} and \eqref{eq:NH}.}
\label{tab:PassiveMech}
\centering
\begin{tabular}{lcccc}
\toprule
\textbf{Domain} & $\mu$ (Pa)& $\alpha$ & $\kappa$ (Pa)& $\rho_0$ (kg/m$^3$)\\
\midrule
$\Omega_\mathrm{V}$ & 325.56 & 22 & $10^6$ & 1082\\
$\Omega_\mathrm{A}$ & 325.56 & 22 & $10^6$ & 1082\\
$\Omega_\mathrm{Valves}$ & $10^6$ & - & $10^6$ & 1082\\
$\Omega_\mathrm{Vessels}$ & $14.9\cdot10^3$ & - & $10^6$ & 1082\\
$\Omega_\mathrm{Peri,apical}$ & $2\cdot10^3$ & - & $10^6$ & 1082\\
$\Omega_\mathrm{Peri,basal}$ & $2\cdot10^3$ & - & $5\cdot10^4$ & 1082\\
$\Omega_\mathrm{Scars}$ & 651.12 & 110 & $10^6$ & 1082\\
\bottomrule
\end{tabular}
\end{table}
The active second Piola-Kirchhoff stress tensor reads
\begin{equation}
    \P_\text{act} = 
                S_\text{a} ( n_f \frac{\vec f_0 \otimes \vec f_0}{\lambda_\mathrm{f}^2}
                + n_s \frac{\vec s_0 \otimes \vec s_0}{\lambda_\mathrm{s}^2}
                + n_n \frac{\vec n_0 \otimes \vec n_0}{\lambda_\mathrm{n}^2} )  \,,
\end{equation}
where $n_f$, $n_s$, and $n_n$ are orthotropic activation parameters in fiber, sheet, and sheet-normal orientation, respectively, and $\lambda$ is the stretch ratio in the given orientation.
Distributing the active stress across these directions can result in a more physiological deformation pattern as shown in \cite{Levrero-Florencio-2019-ID13227} and \cite{Gerach-2020-ID14753}.
In this study however, we choose $n_f = 1$ and $n_s = n_n = 0$ since an increase in stress in the sheet-normal direction restricts LV torsion and stress in the sheet direction results in unphysiological wall thickening and shortening of the LV.
We choose the phenomenological active tension transient $S_\text{a}$ according to \cite{Niederer-2011-ID16420}:
\begin{align}
    &S_\text{a}(\vec x,t,\lambda) = S_\text{peak} \phi (\lambda) \tanh^2(\frac{t_\text{s}}{\tau_c}) \tanh^2(\frac{t_\text{dur} - t_\text{s}}{\tau_r}) \,, \label{eq:Sa} \nonumber\\
    &\phi(\lambda) = \max\qty{\tanh(\text{ld}(\lambda - \lambda_0)),0} \,, \nonumber\\
    &\tau_c = \tau_{c0} + \text{ld}_\text{up} (1 - \phi(\lambda)) \,, \nonumber\\
    &t_\text{s} = t - t_\text{a}(\vec x) - t_\text{emd} \qfor 0 < t_\text{s} < t_\text{dur} \,.
\end{align}
Active tension development is triggered by the local activation time $t_\text{a}(\vec x)$, which is defined as the time when the transmembrane voltage $\V$ reaches a threshold of $\V^\mathrm{thresh} = -20$\,mV.
For a description and values of the remaining parameters, we refer to Table~\ref{tab:ActiveTension}.
%%%%%%
Normally, the process of excitation-contraction-coupling is mediated by an increase of intracellular calcium \citep{keener2009physiology1} immediately after depolarization.
Atrial myocytes typically have shorter action potentials and consequently shorter calcium transients compared to ventricular myocytes due to different gene expressions and ionic channel composition \citep{Ng-2010}.
To account for these differences, we adapted the atrial tension to have a shorter contraction duration such that tension development in the atria and ventricles do not overlap in time.
Furthermore, the model was personalized to match MRI derived data of the subject that serves as the control in this study.
%%%%%%

After discretizing Equation \eqref{eq:linearMomentum} in space using linear tetrahedral elements, the balance of linear momentum equation at time $n+1$ reads
\begin{equation}
    \vec M \vec a^{n+1} + \vec D \vec v^{n+1} + \vec K \vec u^{n+1}  = \vec f_\mathrm{ext}(\vec u^{n+1}, t^{n+1}) \label{eq:discreteMechanics}\,,
\end{equation}
where $\vec M$ is the mass matrix, $\vec K$ is the stiffness matrix, $\vec v$ and $\vec a$ are the first and second derivative  of the displacement $\vec u$ w.r.t time, and $\vec f_\mathrm{ext}$ are external forces.
To represent viscoelastic effects in the material model that arise in the dynamical system \eqref{eq:linearMomentum}, we extend the model by a Rayleigh damping term $\vec D \vec v^{n+1}$ with the damping matrix 
\begin{equation}
    \vec D = \alpha_1 \vec M + \alpha_2 \vec K \,.
\end{equation}
The damping coefficients were set to $\alpha_1 = 500$\,1/s and $\alpha_2 = 0.005$\,s.
Time integration is done implicitly using a Newmark method
\begin{align}
    \vec u^{n+1} &= \vec u^n + \Delta t \vec v^n + \left( \frac{1}{2} - \beta \right) \Delta t^2 \vec a^n \nonumber \\
    & \quad + \beta \Delta t^2 \vec a^{n+1} \,, \label{eq:UpdateDisplacement}\\
    \vec v^{n+1} &= \vec v^n + (1 - \gamma) \Delta t \vec a^n + \Delta t \gamma \vec a^{n+1} \,, \label{eq:UpdateVelocity} \\
    \vec a^{n+1} &= \frac{1}{\beta \Delta t^2} (\vec u^{n+1} - \vec u^n - \Delta t \vec v^n \nonumber \\
    & \quad - \left( \frac{1}{2} - \beta \right) \Delta t^2 \vec a^n) \,, \label{eq:UpdateAcceleration}
\end{align}
with the parameters $\beta = 0.3$, $\gamma = 0.6$, and $\Delta t = 1$\,ms.
Substituting Equations (\ref{eq:UpdateDisplacement},~\ref{eq:UpdateVelocity},~\ref{eq:UpdateAcceleration}) into \eqref{eq:discreteMechanics} results in a nonlinear algebraic equation that only depends on the unknown displacements $\vec u^{n+1}$ which is solved using Newton's method and LU-decomposition implemented in PETSc \citep{Balay2012}.

\subsubsection{Boundary conditions}

\begin{figure*}[htb]
    \includegraphics[width=160mm]{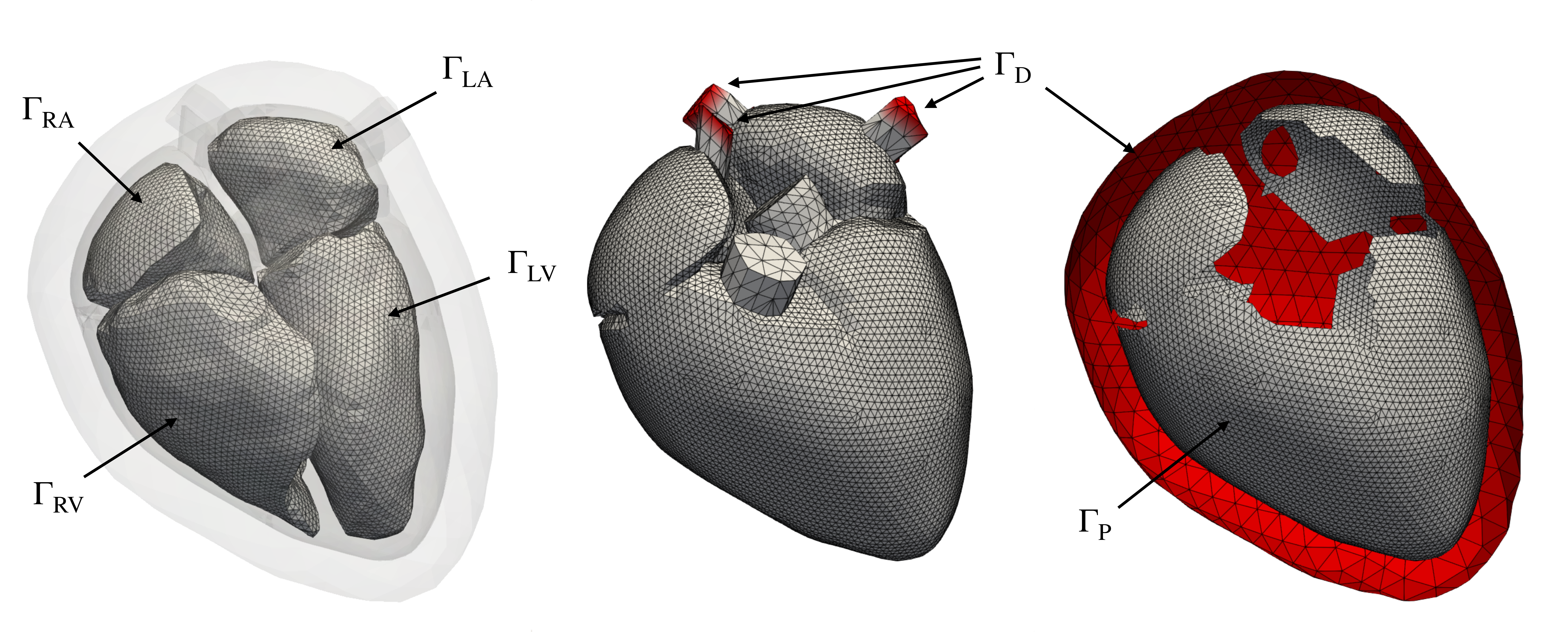}
    \caption{Surfaces on which boundary conditions are applied: the endocardial surfaces $\Gamma_\mathrm{N} = \Gamma_\text{LV}\cup\Gamma_\text{RV}\cup\Gamma_\text{LA}\cup \Gamma_\text{RA}$ are used for the pressure boundary condition; $\Gamma_\mathrm{D}$ is used as a zero displacement Dirichlet boundary; and $\Gamma_\mathrm{P}$ is used for the frictionless sliding contact problem. Note that the Dirichlet boundary in the right figure is clipped for visualization reasons}
    \label{fig:BoundaryConditions}
\end{figure*}

We apply three types of mechanical boundary conditions to the surfaces shown in Figure~\ref{fig:BoundaryConditions}.
Zero displacement Dirichlet boundary conditions
\begin{equation}
    \vec u = \vec 0 \qq{on}\Gamma_\text{D} \times (0,T]	\label{eq:BCDirichlet}
\end{equation} 
are applied to the surfaces defined by $\Gamma_\mathrm{D}$, which comprise the distal ends of the pulmonary veins, the vena cava superior and inferior, and the outer surface of the pericardial tissue.

Additionally, we use the frictionless and permanent contact problem proposed by \cite{fritz13a} that reflects the constraints of the pericardium and the surrounding tissues on the heart:
\begin{equation}
    \F\,\P(\vec x, \F )\vec n - k_\mathrm{epi} g(\vec x) \vec n  = \vec 0 \qq{on} \Gamma_\text{P} \times (0,T]\,. \label{eq:boundcond}
\end{equation}
The contact is established between the surface $\Gamma_\mathrm{P}$ on the epicardium and the surface $\Gamma_\mathrm{S}$, which is defined as the inner surface of the pericardium.
Technically, $\Gamma_\mathrm{P}$ and $\Gamma_\mathrm{S}$ are connected by a spring with spring stiffness $k_\mathrm{epi} = 10$\,MPa through the gap function $g(\vec x)$.
We determine $g(\vec x)$ by projecting a point $\vec x \in \Gamma_\mathrm{P}$ onto the surface $\Gamma_\mathrm{S}$ into the direction of the current normal direction $\vec n$.
Depending on the distance between $\vec x$ and the projected point $\vec x_\mathrm{proj}$, the gap function reads
\begin{equation}
    g(\vec x) = 
    \begin{cases}
    \norm{\vec x - \vec x_\mathrm{proj}}^2 / 2d &\qq*{for} \norm{\vec x - \vec x_\mathrm{proj}} < d \,, \\
    \norm{\vec x - \vec x_\mathrm{proj}} - \frac{d}{2} &\qq*{for} \norm{\vec x - \vec x_\mathrm{proj}} \geq d \,, \\
    0 &\qq*{for} \norm{\vec x - \vec x_\mathrm{proj}} > d_\mathrm{m} \,,
    \end{cases}
\end{equation}
with the transition distance $d = 0.1$\,mm and the maximal distance $d_\mathrm{m} = 8$\,mm below which contact is maintained.
We made this adaptation to the original formulation to yield a smooth transition in case the surfaces $\Gamma_\mathrm{P}$ and $\Gamma_\mathrm{S}$ start to overlap, which results in a change of direction of $\vec n$.
Although we use a fixed spring stiffness $k_\mathrm{epi}$, the spatially varying effect of the surrounding tissue as described by \cite{Strocchi-2020-ID13336} is incorporated into the model by assigning a smaller bulk modulus $\kappa$ to the pericardial tissue in atrial and basal regions of the heart.
Thus, the pericardium becomes more compliant to volumetric changes in these regions and enables a radial contraction of the ventricles around the base without a loss of contact to the surrounding tissue. 

Finally, we use a 0D closed-loop model of the human circulatory system to determine a pressure $p_\mathrm{C}$ for ${\mathrm{C} \in \qty{\mathrm{LV,RV,LA,RA}}}$ in each of the four chambers.
This pressure is applied uniformly on the endocardium as described by the Neumann boundary condition
\begin{equation}
    \F\,\P(\vec x, \F )\vec n - J p_\mathrm{C} \F^{-\mathrm{T}} \vec n = \vec 0 \qq{on} \Gamma_\mathrm{C} \times (0,T] \,. \label{eq:pressurebc}
\end{equation}
The details of this model were recently described in \cite{Gerach-2021}.
Additionally, the model was extended by a valve model based on the description of the instantaneous transvalvular pressure gradient according to \cite{Garcia-2005-ID16440} and a description of valve dynamics according to \cite{Mynard-2012}.
Parameters for the circulatory system are given in Table \ref{tab:CircWholeHeartParameter}.
All four heart chambers of the 3D model are coupled to the 0D model using the condition
\begin{equation}
    \abs{V_\mathrm{C}^\mathrm{3D}(p_\mathrm{C}^{n+1}) - V_\mathrm{C}^\mathrm{0D}(p_\mathrm{C}^{n+1})} < \varepsilon \,,
\end{equation}
with a tolerance $\varepsilon = 10^{-7}$\,mL.

\subsection{Modeling of ablation scars}

\begin{figure*}[htb]
    \includegraphics[width=160mm]{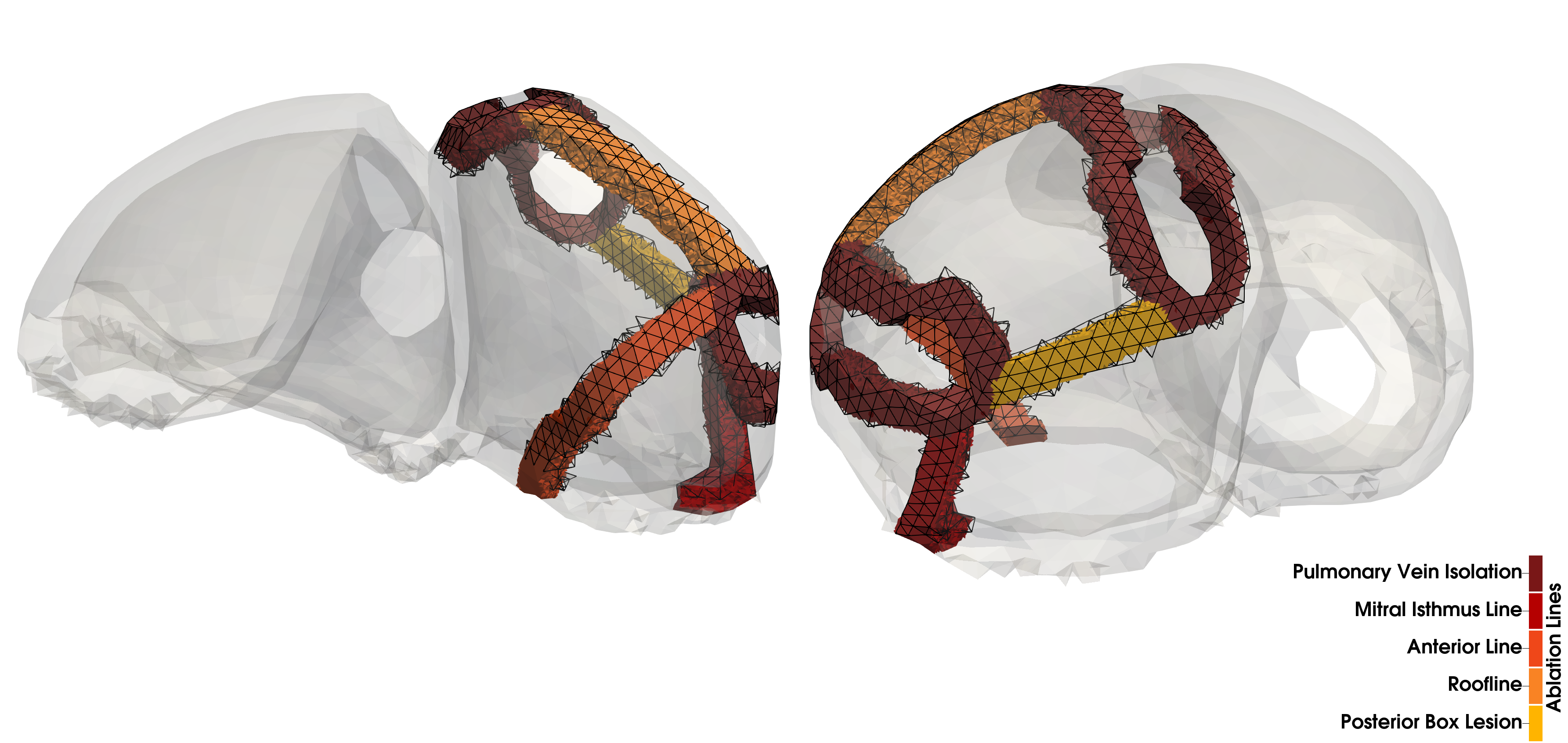}
    \caption{Superior (left) and posterior (right) view of the atria with the considered ablation lesions in the electrophysiological domain $\OmegaEP$: PVI, MIL, AL, RL, BL. The superimposed wiremesh shows the ablation lesions after mapping to the mechanical domain $\OmegaM$.}
    \label{fig:AblationLines}
\end{figure*}

We performed a semi-automatic ablation procedure in the left atrium of our four chamber heart model using the rule-based approach described by \cite{krueger2013patterns,loewe16}.
Five commonly used ablation lesions were placed in the domain $\OmegaEP$: pulmonary vein isolation (PVI), a mitral isthmus line (MIL), an anterior line (AL), a roofline (RL), and a posterior box lesion (BL) which includes the RL.
All lesions were applied transmurally with an average width of 5\,mm and mapped to the mechanical domain $\OmegaM$ (Figure~\ref{fig:AblationLines}).
%%%%%%
Since $\OmegaM$ and $\OmegaEP$ are nested meshes with different resolutions, the mapping error is minimal.
In fact, the ablation lesions cover the same fraction of the atrial myocardium in both meshes.
%%%%%%
For the simulations, we consider the isolated effect of each single ablation lesion as well as all medically feasible combinations.
The electrophysiological tissue conductivity in the affected regions was effectively set to zero to reflect perfect ablations.
Therefore, the threshold voltage $\V^\mathrm{thresh}$ is never reached and no active tension is developed.
Additionally, we model ablated tissue mechanically as isotropic regions with increased tissue stiffness.
Consequently, we adapt the constitutive model in Equation~\eqref{eq:usyk} by setting the anisotropic scaling parameters to $b_\mathrm{ff} = b_\mathrm{ss} = b_\mathrm{nn} = 1$ and $b_\mathrm{fs} = b_\mathrm{fn} = b_\mathrm{ns} = 0.5$.
The increased stiffness of the scar tissue is achieved by choosing $\alpha$ five times higher and $\mu$ two times higher than in the bulk tissue \citep{Niederer-2011-ID16420}.
This approximates the expected increase in tissue stiffness due to a build-up of collagen in the myocardium as observed in animal studies \citep{Jugdutt-1996-ID16489}. The sensitivity of the results regarding the scar stiffness is evaluated in an additional experiment.

\subsection{Simulation setup and initialization}
\label{sec:init}

In total, we conduct ten simulations including nine combinations of the five ablation lines shown in Figure~\ref{fig:AblationLines} and a control case without ablation.

First, the cellular models are paced at a cycle length of 1.2\,s for a total of 1000 cycles.
The final values of the state variables $\qty{\V, \vec w, \vec c}$ are used as initial values on the vertices of $\OmegaEP$.
Second, the reference configuration $\OmegaM$ was unloaded using a backward displacement method \citep{Marx-2021-ID15684}.
We used typical diastatic pressure values in the four chambers: $p_\mathrm{RV} = p_\mathrm{RA} = 4$\,mmHg; $p_\mathrm{LV} = p_\mathrm{LA} = 8$\,mmHg.
Subsequently, the unloaded configuration was inflated with the same pressure values to pre-stress the tissue.
Even though the ablation lines have different material parameters than healthy tissue, we used the same unloaded configuration for all ten simulation setups.
As long as the global stiffness of the material is similar, this assumption should not affect the overall results in this study \citep{Kovacheva-0000-ID14277}.

A comparison of the model outputs is only relevant if the system has reached a stable limit cycle.
In the four chamber model, this means that the blood volume distribution does not change anymore during subsequent cardiac cycles.
Estimating the time it takes for the model to reach this state is difficult, hence we implement an automated stopping criterion based on the stroke volume difference $\mathrm{SV_{diff}}$ between the LV and the RV by solving an additional differential equation in the 0D circulation model:
\begin{equation}
   \partial_t \mathrm{SV_{diff}} = Q_\mathrm{SysArt} - Q_\mathrm{PulArt}\,,
\end{equation}
where $Q_\mathrm{SysArt}$ and $Q_\mathrm{PulArt}$ is the flow through the aorta and the pulmonary artery, respectively.
At integer multiples of the cycle length, $\abs{\mathrm{SV_{diff}}}$ is compared to a pre-defined threshold (1\,ml) and reset to zero if the criterion is not fulfilled yet.

\section{Results}
\label{results}

We investigated the influence of standard clinical ablation patterns on the pumping efficiency of the human heart by simulating a total of ten cases:
First, we calibrated the model parameters to match the cine MRI data of the volunteer to simulate a healthy control case.
Next, we introduced different combinations of the basic ablation patterns shown in Figure~\ref{fig:AblationLines} into the left atrium in nine additional simulations.
All ten simulations were analyzed with regards to the activation sequence, blood volume, and correlation between the amount of inactive and ablated tissue with ejection fraction (EF) of the left atrium.
Furthermore, we investigated the influence of the ablation lesions on the ventricular function with a focus on AVPD and ventricular stroke volume (SV).

\begin{figure*}[htb]
    \centering
    \begin{subfigure}[b]{0.32\textwidth}
        \includegraphics[width=\textwidth]{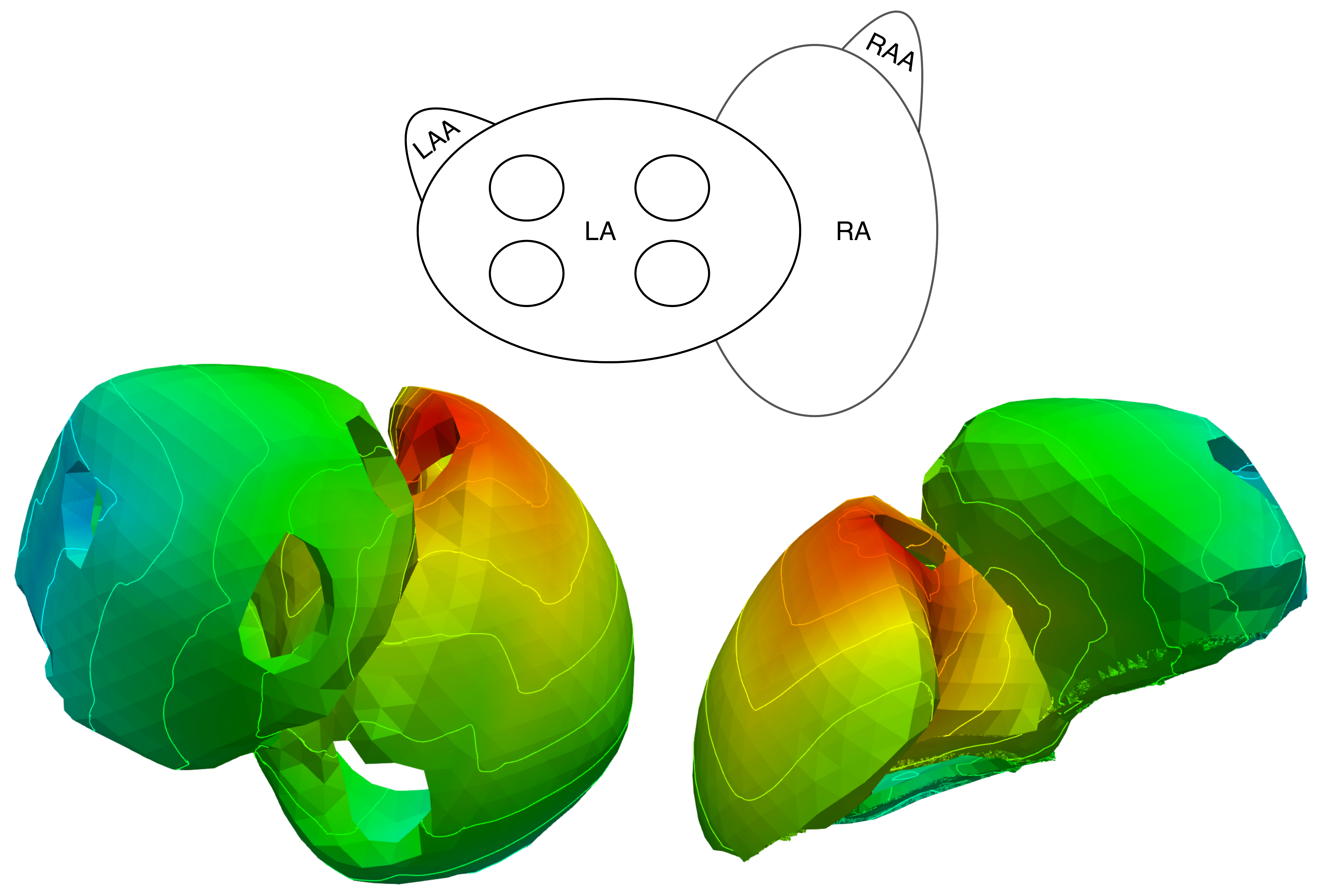}
        \caption{Control}
        \label{fig:LATHealthy}
    \end{subfigure}
    \begin{subfigure}[b]{0.32\textwidth}
        \includegraphics[width=\textwidth]{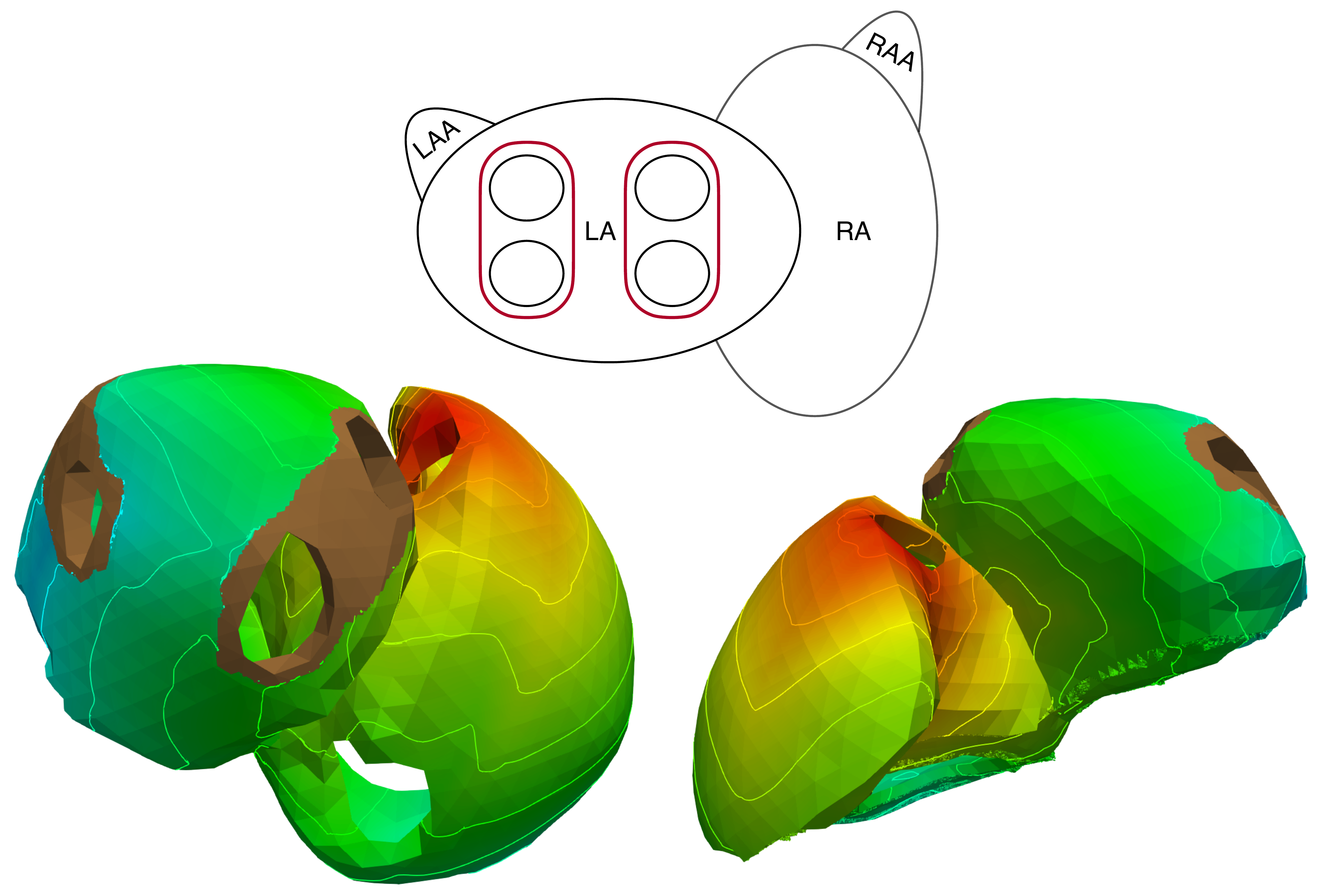}
        \caption{PVI}
    \end{subfigure}
    \begin{subfigure}[b]{0.32\textwidth}
        \includegraphics[width=\textwidth]{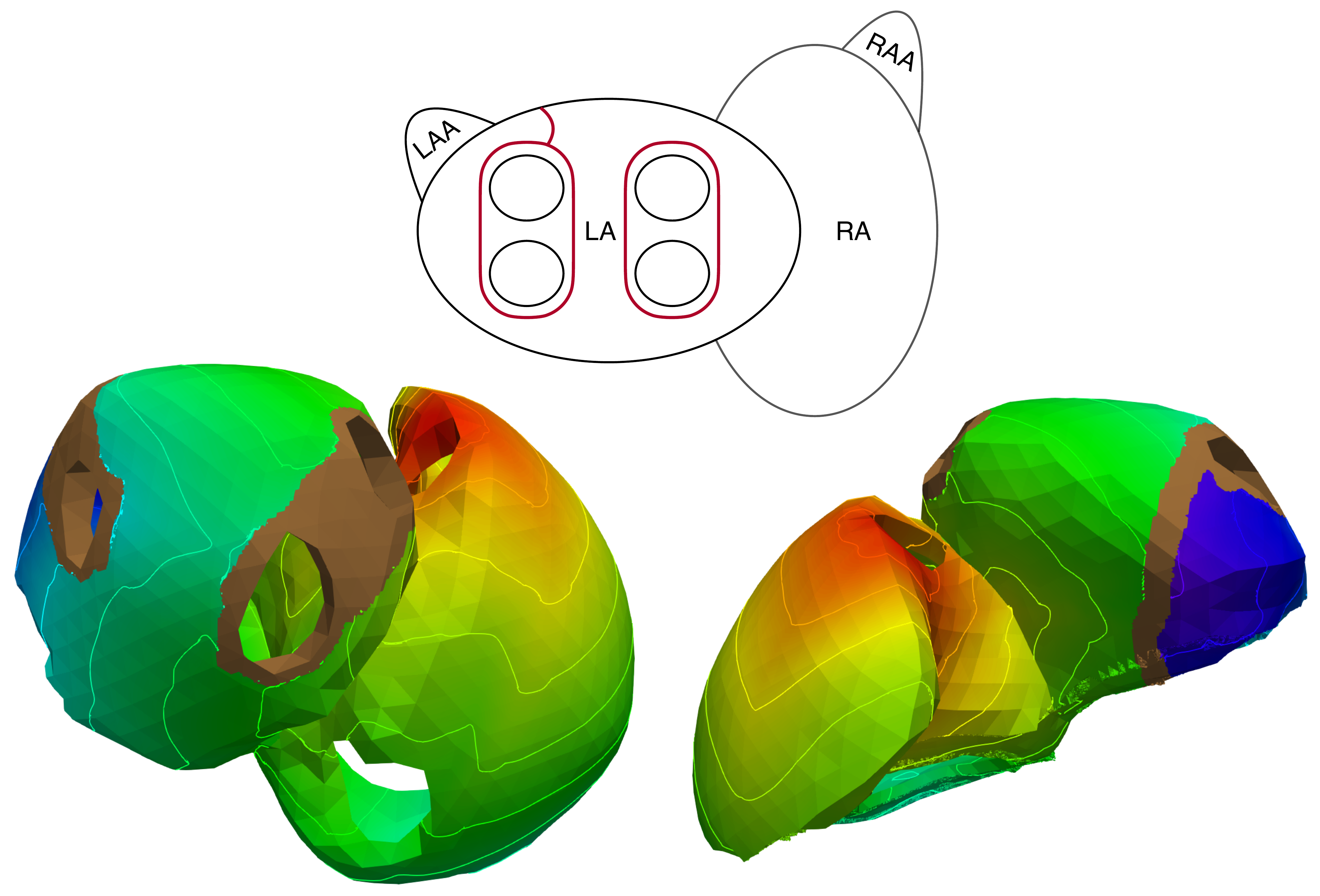}
        \caption{PVI+AL}
    \end{subfigure}
    \begin{subfigure}[b]{0.32\textwidth}
        \includegraphics[width=\textwidth]{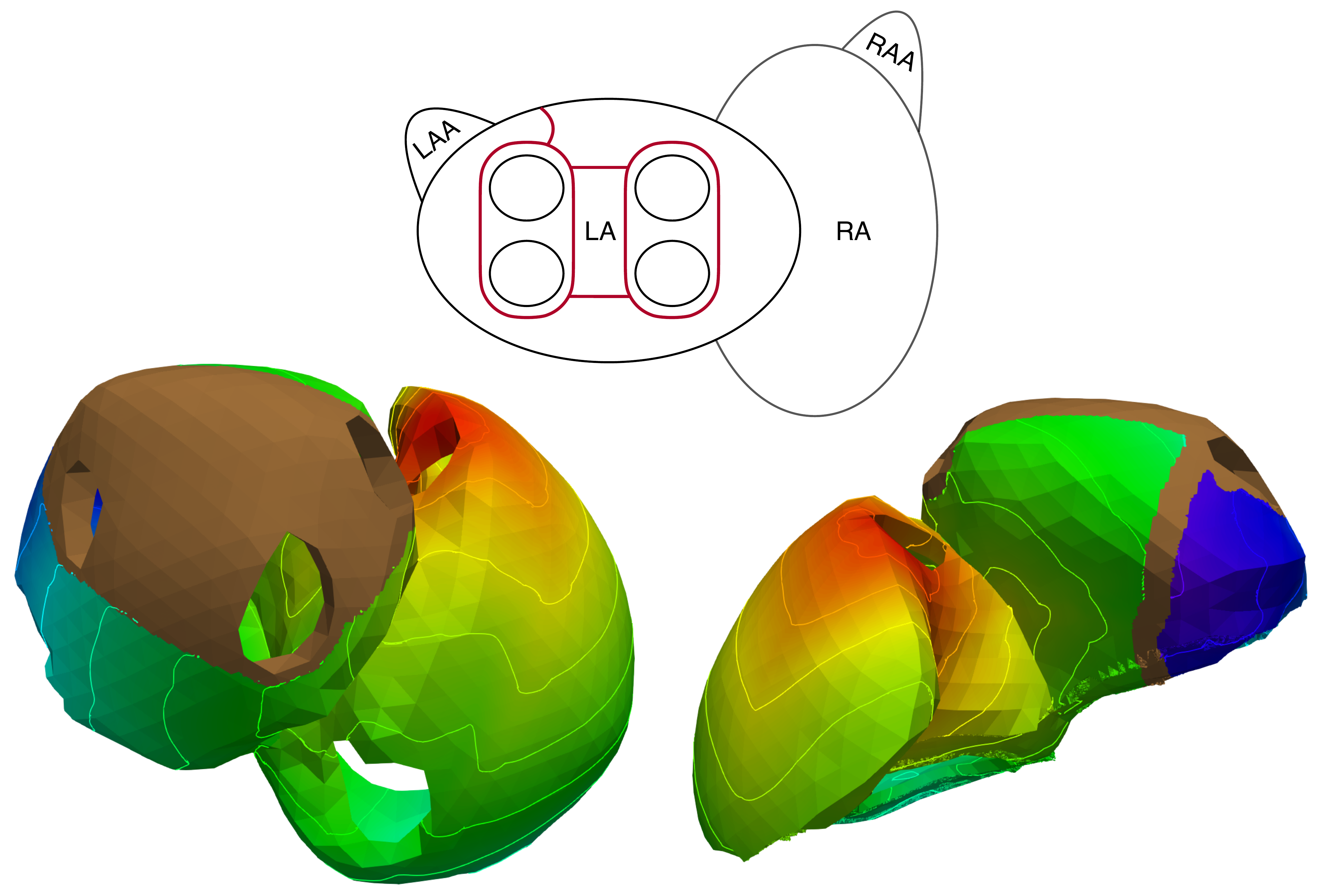}
        \caption{PVI+AL+BL}
    \end{subfigure}
    \begin{subfigure}[b]{0.32\textwidth}
        \includegraphics[width=\textwidth]{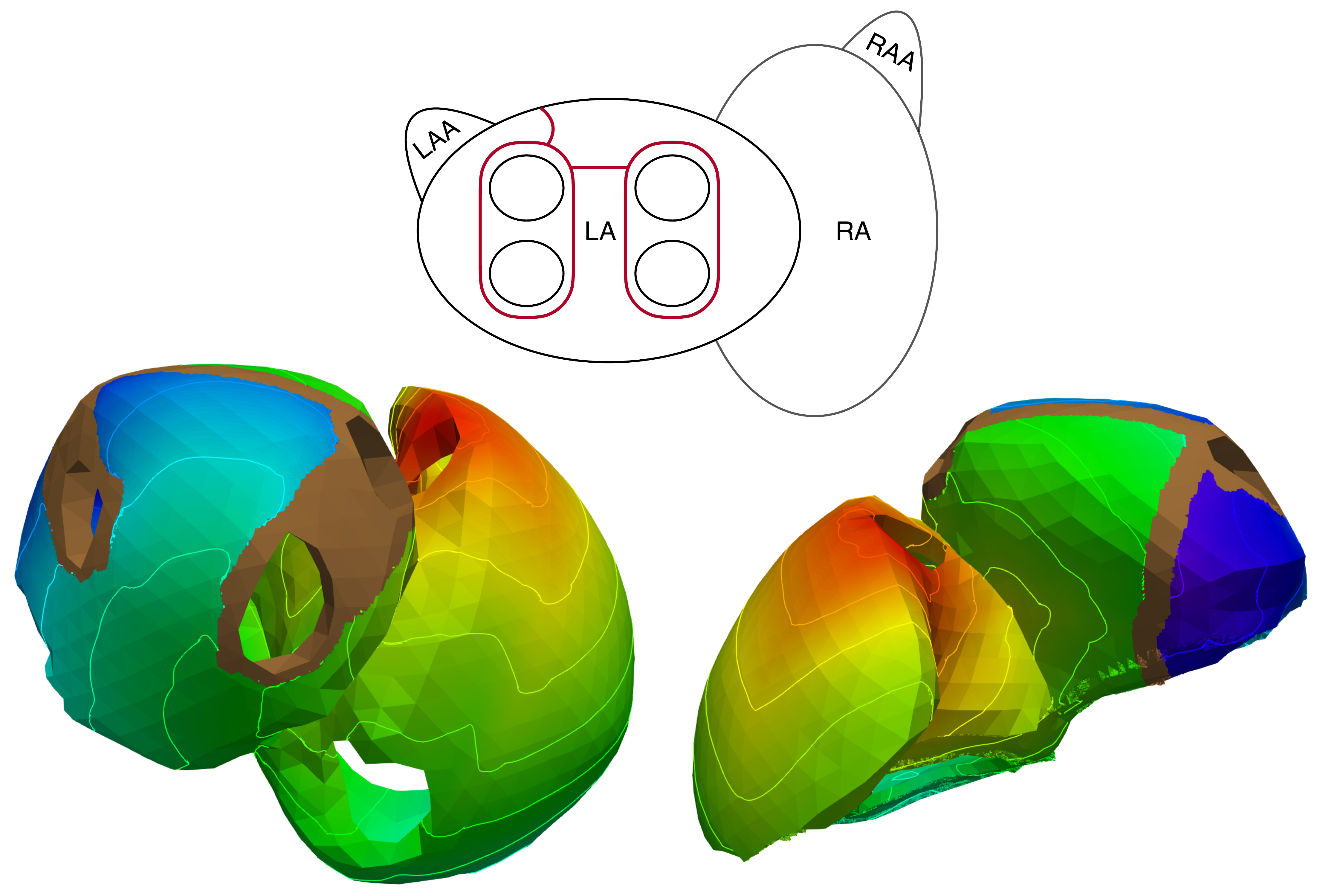}
        \caption{PVI+AL+RL}
    \end{subfigure}
    \begin{subfigure}[b]{0.32\textwidth}
        \includegraphics[width=\textwidth]{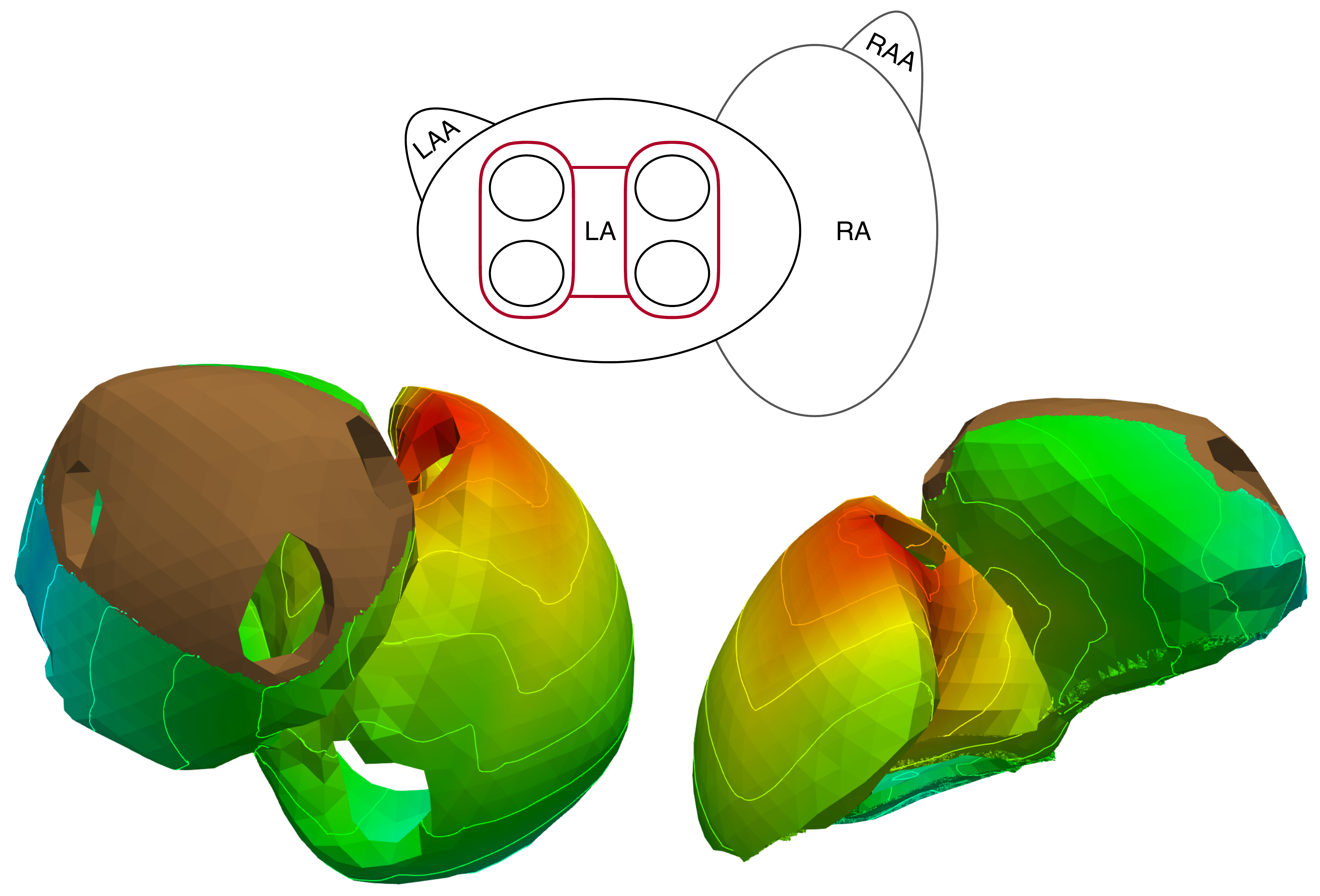}
        \caption{PVI+BL}
    \end{subfigure}
    \begin{subfigure}[b]{0.32\textwidth}
        \includegraphics[width=\textwidth]{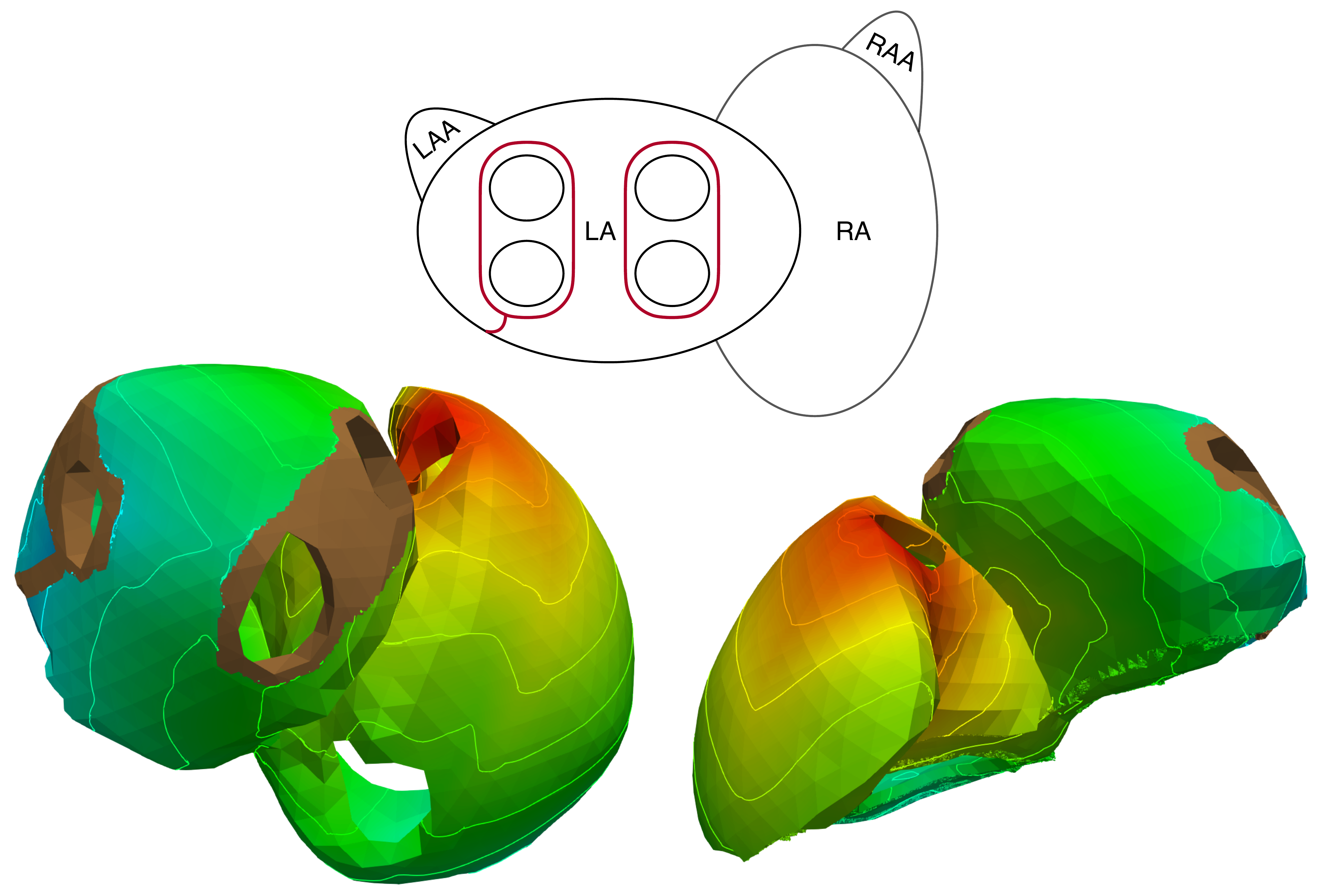}
        \caption{PVI+MIL}
    \end{subfigure}
    \begin{subfigure}[b]{0.32\textwidth}
        \includegraphics[width=\textwidth]{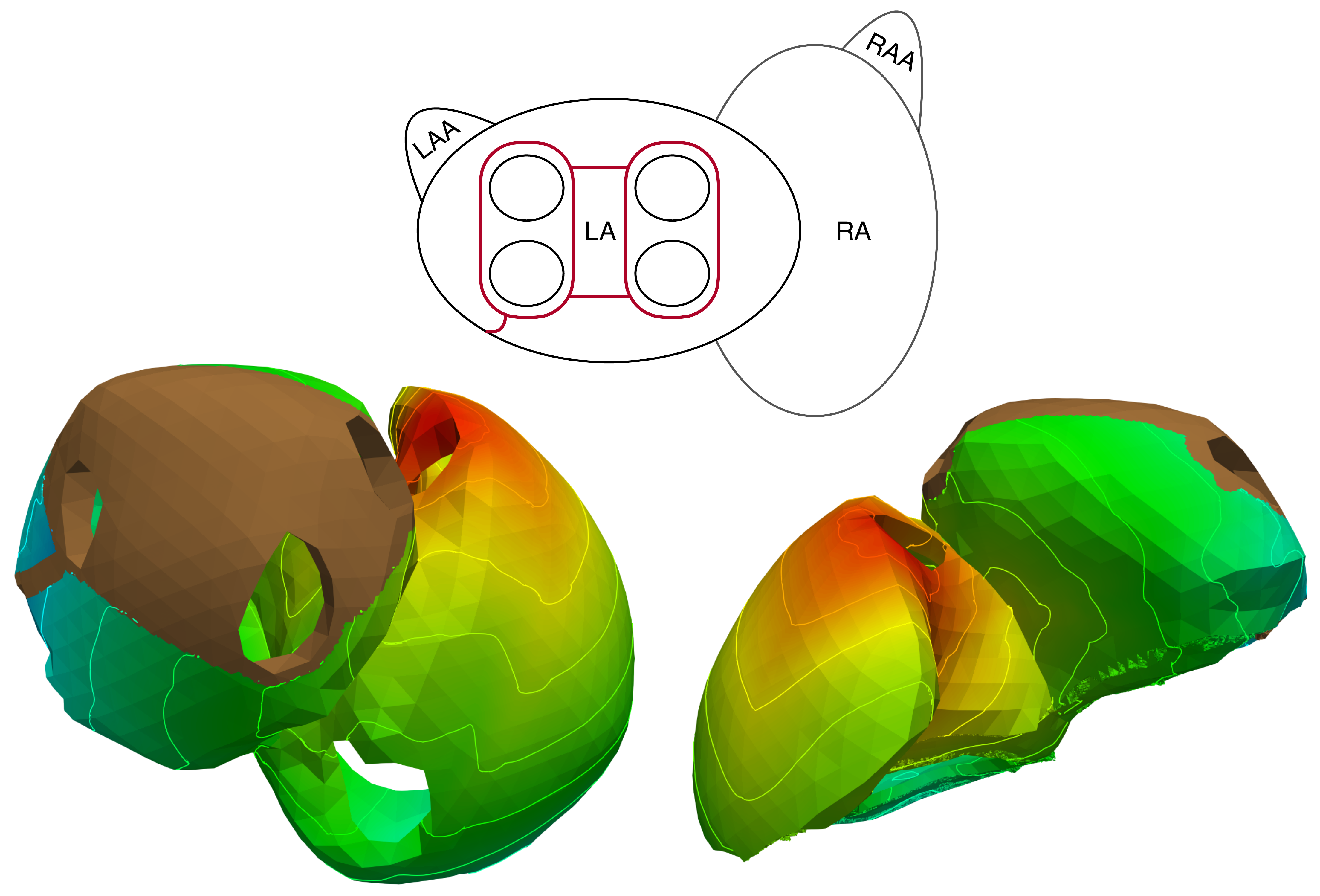}
        \caption{PVI+MIL+BL}
    \end{subfigure}
    \begin{subfigure}[b]{0.32\textwidth}
        \includegraphics[width=\textwidth]{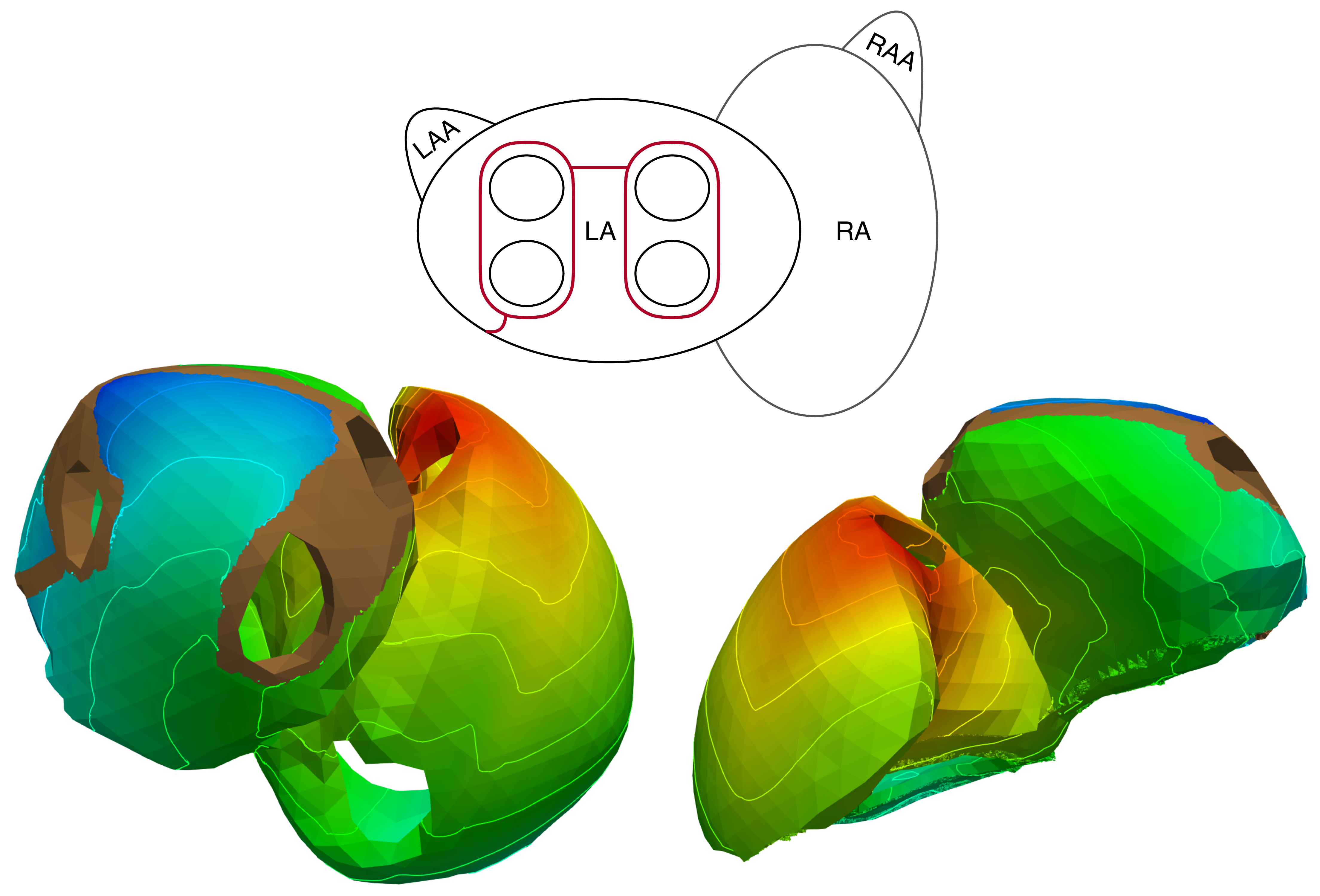}
        \caption{PVI+MIL+RL}
    \end{subfigure}
    \begin{subfigure}[b]{0.32\textwidth}
        \includegraphics[width=\textwidth]{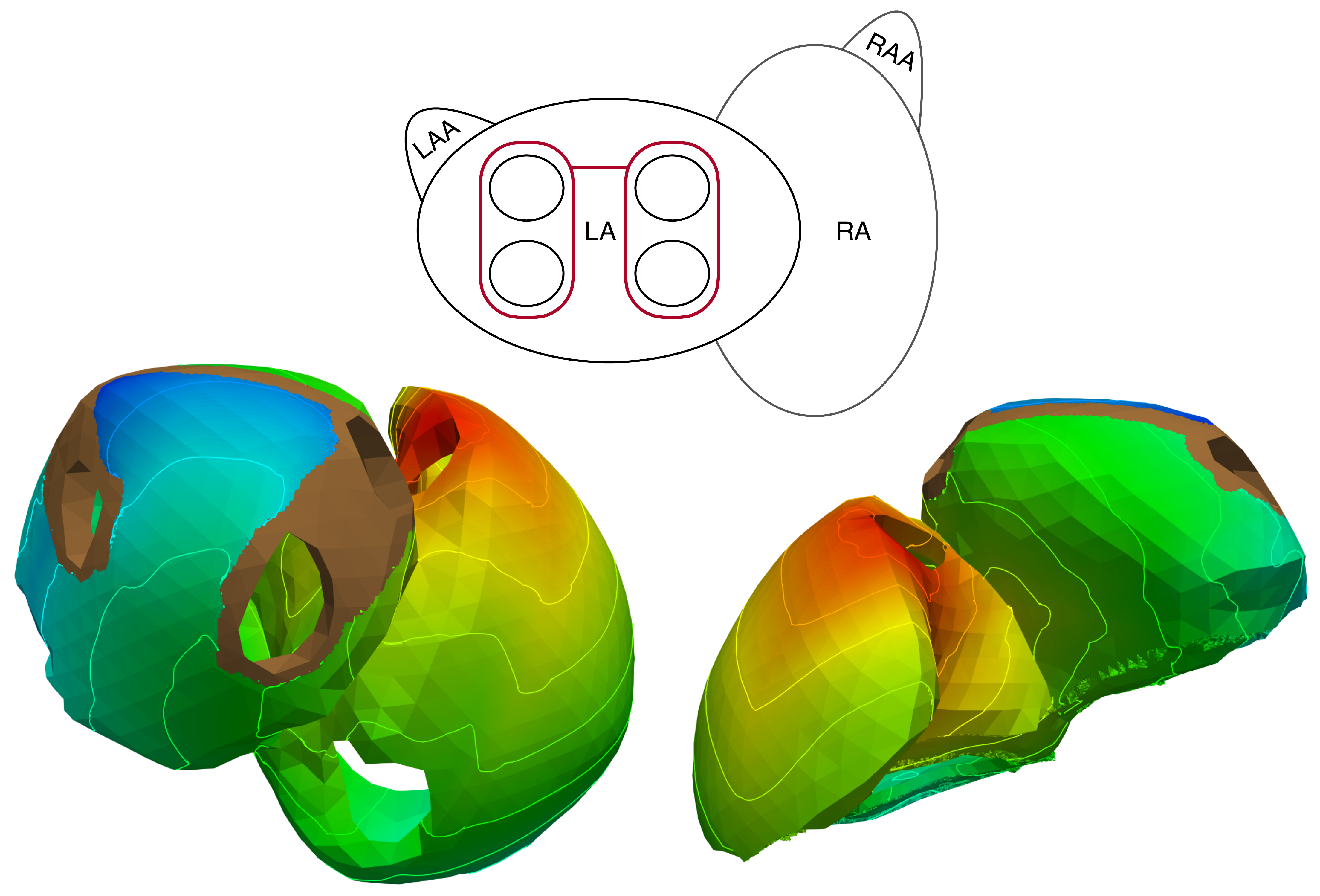}
        \caption{PVI+RL}
    \end{subfigure}
    \begin{subfigure}[b]{0.1\textwidth}
        \includegraphics[width=\textwidth]{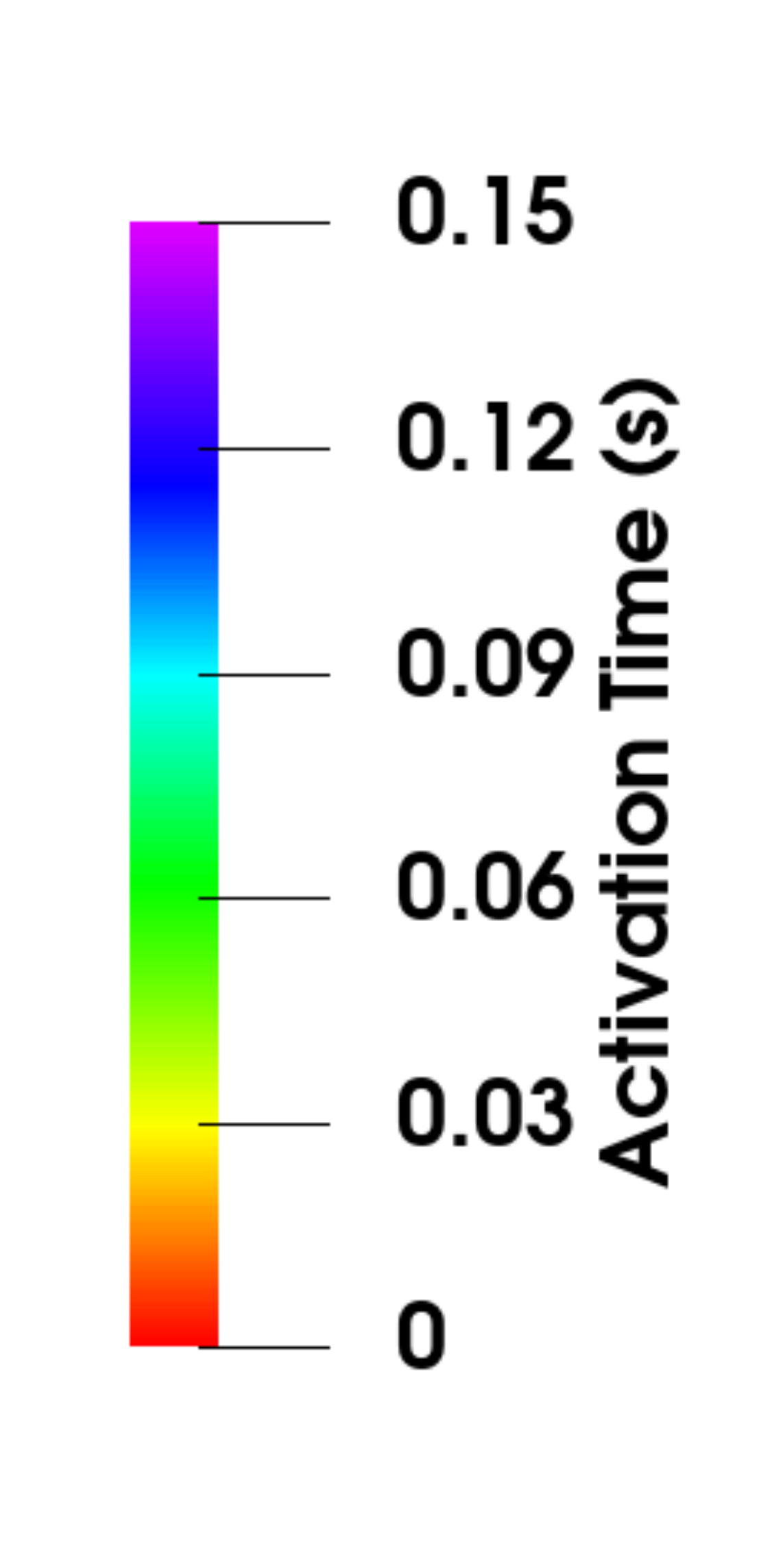}
    \end{subfigure}
    \caption{Activation maps and isochrones of the depolarization wave in sinus rhythm. A brown color denotes inactive tissue. The pictogram of the atria at the top of each model indicates the applied ablation lesions (red lines).}
    \label{fig:LAT}
\end{figure*}

Figure~\ref{fig:LAT} shows the atrial activation time maps for all ten simulations during sinus rhythm.
In the healthy control case (Figure~\ref{fig:LATHealthy}), the activation starts at the sinus node and propagates from the right atrium (RA) to the left atrium (LA) via four pathways: the Bachmann bundle on the anterior side, a middle and upper posterior interatrial connection, and via the coronary sinus.
The depolarization in the left atrium starts after 29\,ms at the entry site of the Bachmann bundle.
The last activation of the atria occurs after 97\,ms at the posterior side of the mitral valve annulus.
The ablation lesions that have the biggest influence on atrial activation time are PVI+RL and PVI+AL.
As a consequence of the RL lesion, the activation of the left atrial roof can only occur from the posterior side resulting in an activation delay of around 30\,ms in this area ($\approx 100$\,ms in PVI+AL+RL compared to $\approx 70$\,ms in Control).
The AL lesion blocks an activation of the left atrial appendage (LAA) through the Bachmann bundle on the anterior side, thus delaying the depolarization of the LAA significantly.
Therefore, total atrial activation takes 147\,ms in cases with an AL lesion.
Compared to PVI+AL, PVI+MIL does not alter the activation sequence significantly, the MIL lesion coincides with the area of latest activation in the LA in our setup.
Circumferential PVI alone does not notably alter the activation sequence of the LA.
Although the activation sequence is not drastically changed by PVI+BL, a large amount of tissue stays inactive due to the isolation of the LA posterior wall by this specific scar.

\begin{figure*}
    \centering
    \includegraphics[width=125mm]{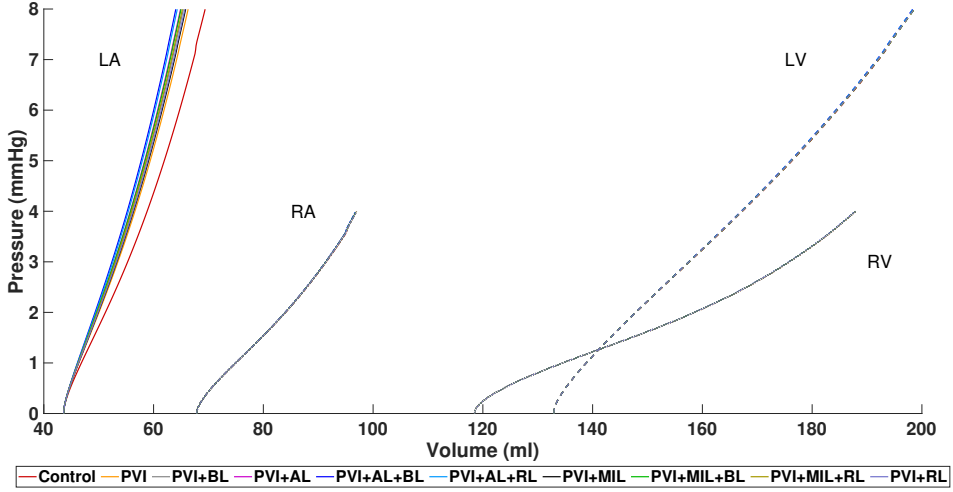}
    \caption{End-diastolic pressure-volume relationship (EDPVR) for the left atrium (LA), right atrium (RA), left ventricle (LV) and right ventricle (RV) for all simulations.}
    \label{fig:EDPVR}
\end{figure*}

Figure~\ref{fig:EDPVR} shows the end-diastolic pressure-volume relationship (EDPVR) of all four chambers resulting from the inflation of the pressure-free state during the initialization process described in Section~\ref{sec:init}.
Starting from an identical pressure-free state, we observe that the volume of the LA in the pre-stressed state is reduced by up to 5.4\,mL ($-$7.8\%) if stiffer scar tissue is present compared to the healthy case.
However, the reduction in volume does not correlate with the amount of ablated tissue in the LA (e.g., PVI+AL has a lower volume and less ablated tissue than PVI+BL).
The volume of the other chambers remains almost identical with differences of less than $\pm0.1\%$.

\begin{figure*}
    \centering
    \includegraphics[width=125mm]{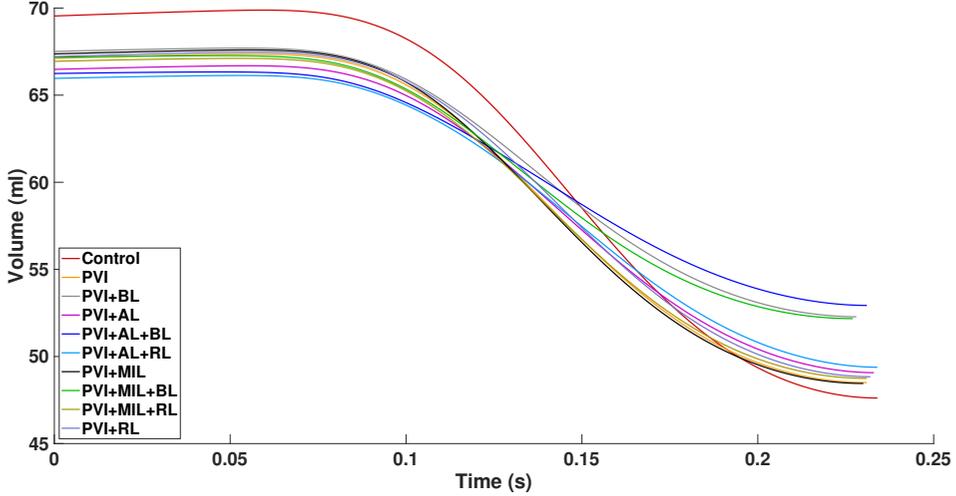}
    \caption{Blood volume traces during left atrial systole. Atrial systole lasts from the externally applied stimulation of the sinus node to the time of maximal contraction of the left atrium.}
    \label{fig:LASystole}
\end{figure*}

Figure~\ref{fig:LASystole} shows the amount of blood in the LA until peak LA systole.
Peak systole was determined as the time when the maximal output of blood volume is reached.
We can observe for all cases that it takes a slightly different amount of time to reach this point.

\begin{figure*}
\centering
    \includegraphics[width=125mm]{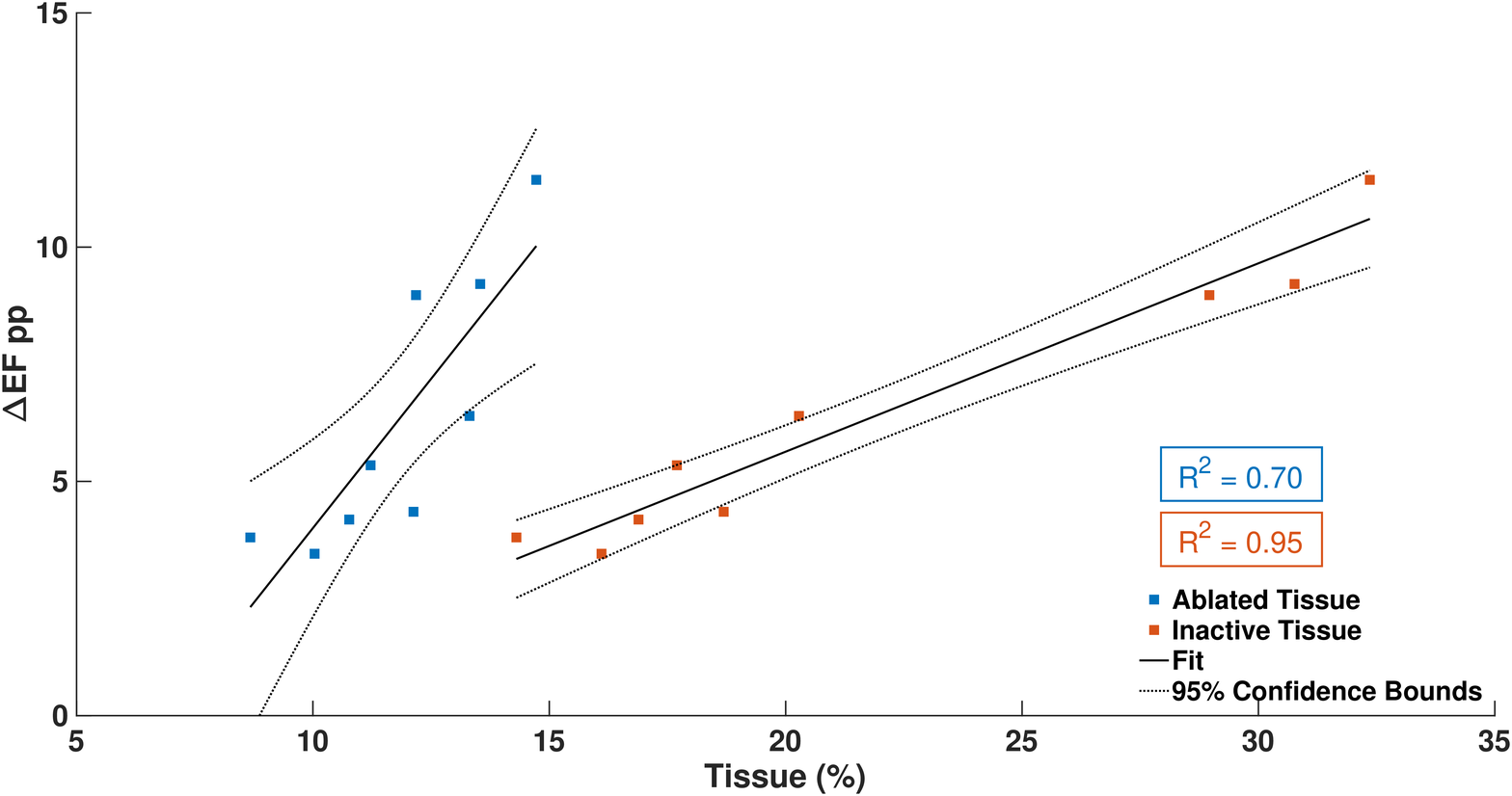}
    \caption{Relationship between the ablated tissue (blue squares)/inactive tissue (orange squares) and the percentage point difference in ejection fraction ($\Delta$EP pp) with a fitted linear regression model.}
    \label{fig:correlation}
\end{figure*}

The change in atrial EF due to the different ablation patterns is summarized in Table~\ref{tab:LASystole} alongside the amount of ablated and subsequently inactive tissue.
A linear regression model shows that the amount of inactive tissue is highly correlated ($R^2 = 0.95$) with the change in ejection fraction ($\Delta$EF) compared to the healthy case (Figure~\ref{fig:correlation}).
Ablated tissue and $\Delta$EF show a much weaker correlation ($R^2 = 0.70$).

\begin{table*}[htb]
\caption{Indicators of left atrial function for the control case and different ablation lesions. SV: stroke volume, EF: ejection fraction, AT: ablated tissue, IT: inactive tissue.}
\label{tab:LASystole}
\centering
\begin{tabular}{lrrrrr}
\toprule
\textbf{Case} & \textbf{SV (mL)} & \textbf{EF (\%)} & \textbf{$\Delta$EF (pp)} & \textbf{AT (\%)} & \textbf{IT (\%)} \\
\midrule
Control & 21.93 & 31.53 & 0 & 0 & 0 \\
PVI & 18.61&	27.73&	3.80&	8.67&	14.30 \\
PVI+BL&	15.22&	22.55&	8.97&	12.18&	28.96 \\
PVI+AL&	17.41&	26.18&	5.34&	11.22&	17.69 \\
PVI+AL+BL&	13.31&	20.09&	11.43&	14.72&	32.35 \\
PVI+AL+RL&	16.58&	25.13&	6.39&	13.31&	20.28 \\
PVI+MIL&	18.91&	28.07&	3.45&	10.03&	16.10 \\
PVI+MIL+BL&	14.99&	22.31&	9.21&	13.53&	30.76 \\
PVI+MIL+RL&	18.19&	27.18&	4.35&	12.12&	18.68 \\
PVI+RL&	18.38&	27.34&	4.18&	10.76&	16.88 \\
\bottomrule
\textbf{pp} percentage points & & & & &
\end{tabular}
\end{table*}

\begin{figure*}[htb]
    \centering
    \includegraphics[width=160mm]{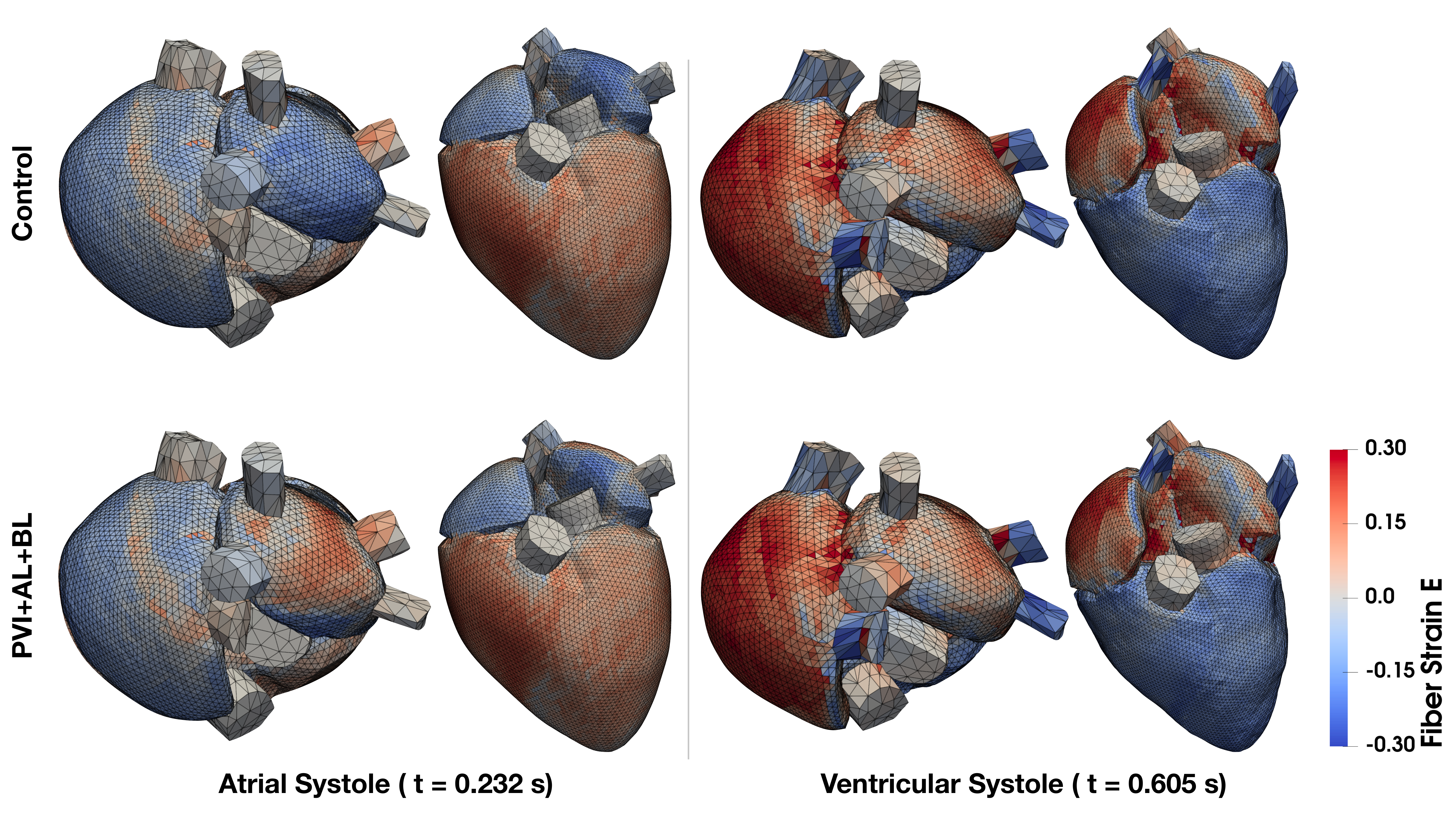}
    \caption{Green Lagrange strain $E_\mathrm{ff}$ in fiber direction of the Control and PVI+AL+BL case is shown during atrial and ventricular systole, respectively.}
    \label{fig:strain}
\end{figure*}

%%%%%%
The strain maps of the Control and PVI+AL+BL cases are shown in Figure~\ref{fig:strain}.
During atrial systole, two differences can be observed in the PVI+AL+BL case compared to the Control case.
First, strain values at the position of the ablation lesions suggest that the fibers remain elongated during atrial contraction.
However, they shorten a bit compared to the diastolic state.
Second, the isolated posterior roof of the left atrium gets stretched past the strain values during diastole since there is no contraction of the fibers in that area.
During ventricular systole, the differences in the strain maps between the Control and PVI+AL+BL are very small.
Only in the scars themselves a 50\% reduced fiber strain can be observed.
The increased stiffness of the scars has a more pronounced effect on the mean stress $\P_\mathrm{ff} = \vec f_0 \cdot \P \vec f_0$ of the tissue.
In general, mean diastolic stress is increased two to three times in the posterior roof and scar surroundings (3.7\,kPa vs. 1.2\,kPa).
Due to the missing contraction of the posterior wall in the PVI+AL+BL case, mean stress is generally lower during atrial systole.
During ventricular systole however, mean stress is increased two-fold in scar tissue (42\,kPa vs. 18\,kPa) while stress in the posterior wall is nearly equal between the two cases. 
%%%%%%

\begin{figure*}[htb]
    \centering
    \includegraphics[width=160mm]{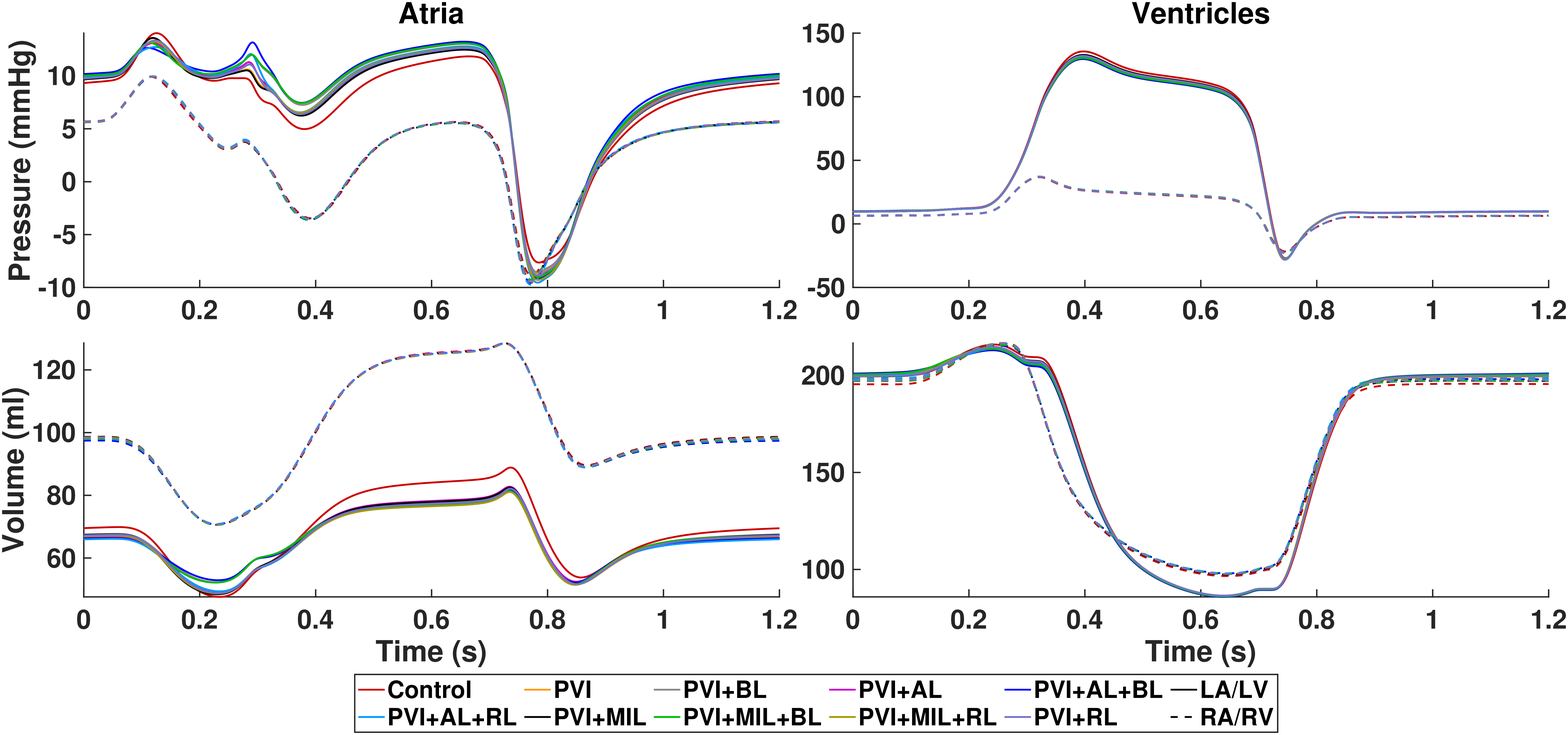}
    \caption{Atrial (left column) and ventricular (right column) pressure (top row) and volume (bottom row) courses during one heart cycle. Different colors denote the simulation results of the healthy control case and all simulations including scars. Left atrium and ventricle are shown with solid lines while right atrium and ventricle are shown with dashed lines.}
    \label{fig:chambers}
\end{figure*}

Figure~\ref{fig:chambers} shows the time course of pressure and volume in all four chambers of the heart.
Besides the decrease in systolic function of the LA, we observe that passive filling of the LA during ventricular contraction is restricted (7\% to 10\% lower compared to control) in the presence of ablation lesions.
At the same time, we observe an increase in LA pressure of up to 12\% compared to the control case.
Furthermore, there is an additional peak in LA pressure at the beginning of ventricular systole (0.28\,s) which is more pronounced in simulations that include the ablation lesions AL and BL, suggesting the changed activation sequence or the increase in inactive tissue as a source.
In the case of PVI+AL+BL, the pressure is 20\% higher compared to the control case.
Except for small deviations in volume (within 1\,mL of each other) at the beginning of the heart beat, the RA is unaffected by the ablation lesions in the LA.

\begin{figure*}[htb]
    \centering
    \includegraphics[width=160mm]{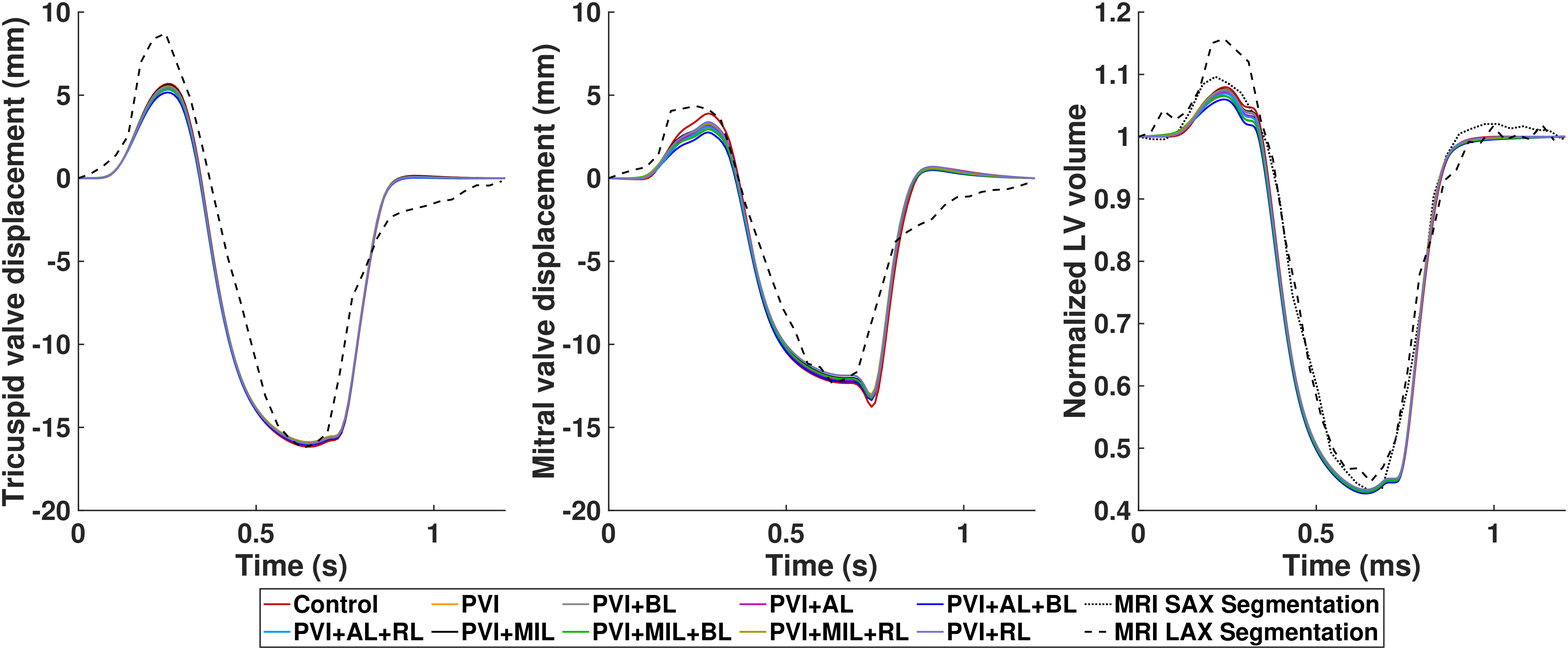}
    \caption{Left: Atrioventricular valve plane displacement (AVPD) of the tricuspid valve. Center: AVPD of the mitral valve. Right: Normalized blood volume in the left ventricle during one heart cycle. Different colors denote the simulation results of the healthy control case and all simulations including scars. Simulation results are compared to data from cine MRI short axis (SAX, dotted line) and long axis (LAX, dashed line) segmentations.}
    \label{fig:AVPD}
\end{figure*}

The contribution of the atrial contraction to ventricular filling reduces in the presence of scars in the LA.
This can be observed in the reduced end-diastolic volume (EDV) of the LV.
However, this effect is not very pronounced and only reduces the volume contribution by the atria by up to 1.4\% as shown in Figure~\ref{fig:AVPD} (right graph).
End-systolic volumes (ESV) of the LV lie within 1\% of each other while peak systolic pressure is reduced by 6\,mmHg.
Furthermore, the increased stiffness of the scars in the LA caused a reduction of the AVPD of the mitral valve (Figure~\ref{fig:AVPD}, middle graph) from +3.90\,mm in the control case to +2.76\,mm in PVI+AL+BL during atrial contraction.
During ventricular contraction, AVPD is reduced from $-$12.33\,mm in the control case to $-$11.87\,mm in PVI+MIL+RL and PVI+RL.
Similar behavior for AVPD and blood volume can be observed in the RV (Figure~\ref{fig:chambers} and \ref{fig:AVPD}, left graph).

%%%%%%
\subsection{Model sensitivity towards scar stiffness}

\begin{figure}[htb]
    \centering
    \includegraphics[width=75mm]{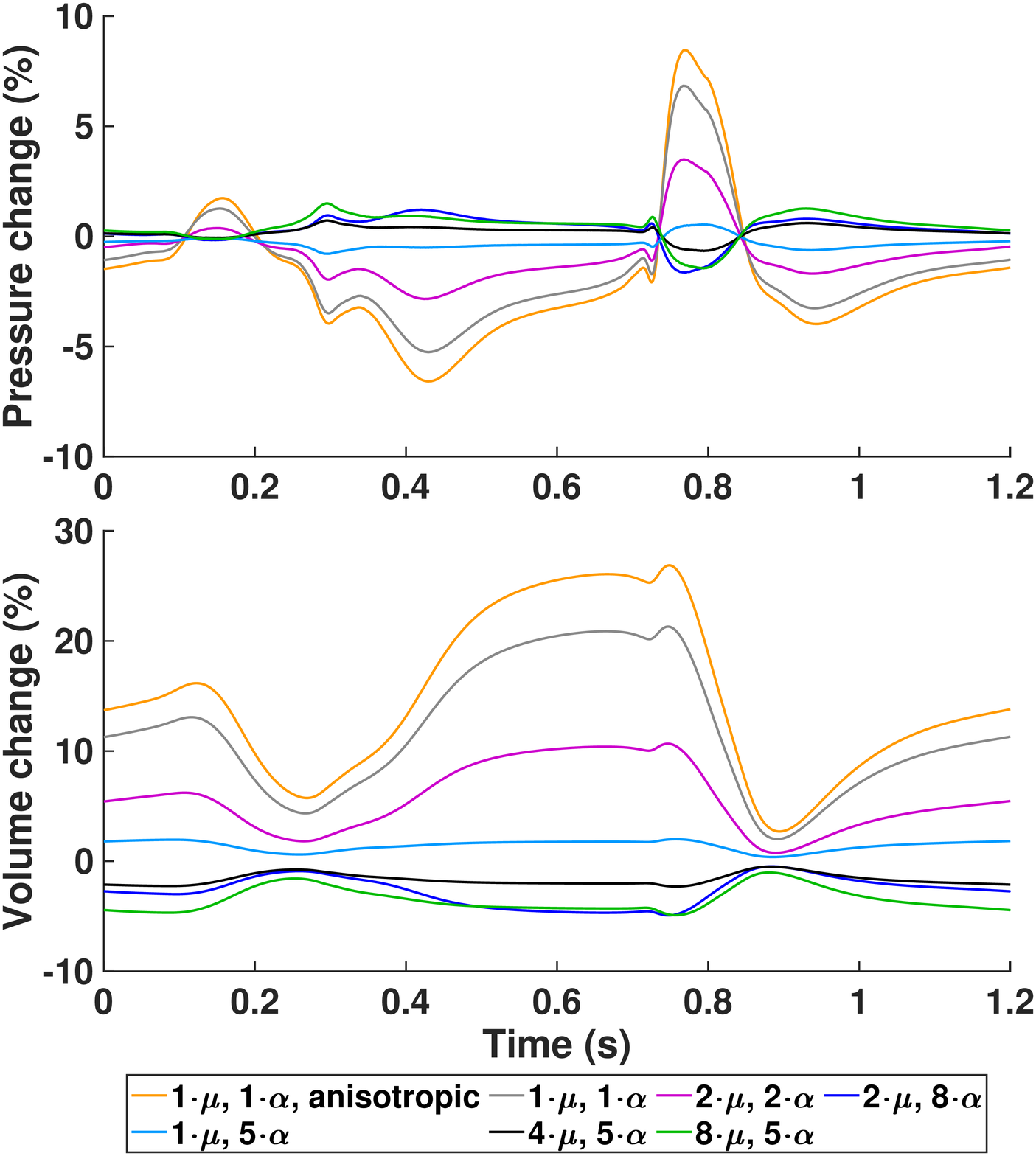}
    \caption{Simulation results of the case PVI+AL+BL for varying scar stiffness. The change in pressure and volume of the left atrium is relative to a PVI+AL+BL reference simulation with $2\cdot \mu$ and $5\cdot \alpha$ and normalized by their peak-to-peak values.}
    \label{fig:scarSens}
\end{figure}

Since there is not a lot of definitive data on how the stiffness of atrial scar tissue changes compared to healthy myocardium, a sensitivity analysis was conducted to investigate how scar stiffness relates to pressure and volume changes in the heart.
Stiffness is mainly controlled by the parameters $\mu$ and $\alpha$ in the constitutive model \eqref{eq:usyk}.
Therefore, different combinations of these parameters were chosen for the scars in the PVI+AL+BL case and the outcome on pressure and volume was evaluated based on the relative difference $x_\mathrm{diff}$ to a PVI+AL+BL simulation with $2\cdot \mu$ and $5\cdot \alpha$:
\begin{equation*}
x_\mathrm{diff} = \frac{x - x^\mathrm{ref}}{x_\mathrm{max}^\mathrm{ref} - x_\mathrm{min}^\mathrm{ref}} \,.
\end{equation*}
The results for the left atrium are shown in Figure~\ref{fig:scarSens}.
The biggest difference was observed when scar tissue is modeled using the same parameters as for healthy myocardium ($1\cdot \mu$ and $1\cdot \alpha$, anisotropic).
Both, pressure and volume differences, are largest during ventricular contraction and relaxation.
Pressure changed by up to 9\% and volume by up to 25\%.
In general, an increase in stiffness is related to a decrease in volume and an increase in pressure (vice versa with decreased stiffness).
Changes in the other chambers of the heart are in the range of 1\% to 3\%.

\subsection{Reduction of conduction velocity}

\begin{figure*}[htb]
    \centering
    \begin{subfigure}[b]{0.32\textwidth}
        \includegraphics[width=\textwidth]{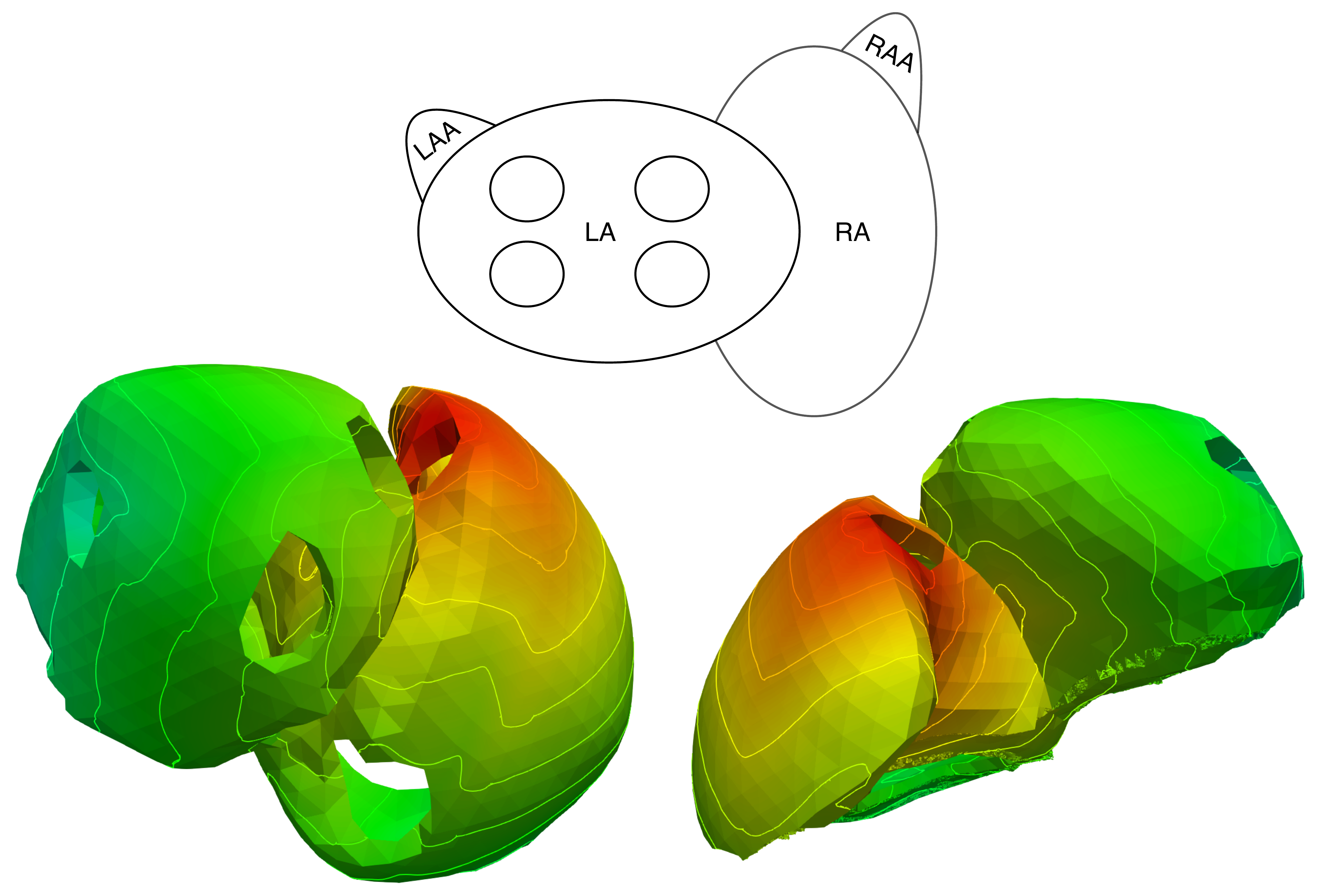}
        \caption{Control, 25\% red.}
    \end{subfigure}
    \begin{subfigure}[b]{0.32\textwidth}
        \includegraphics[width=\textwidth]{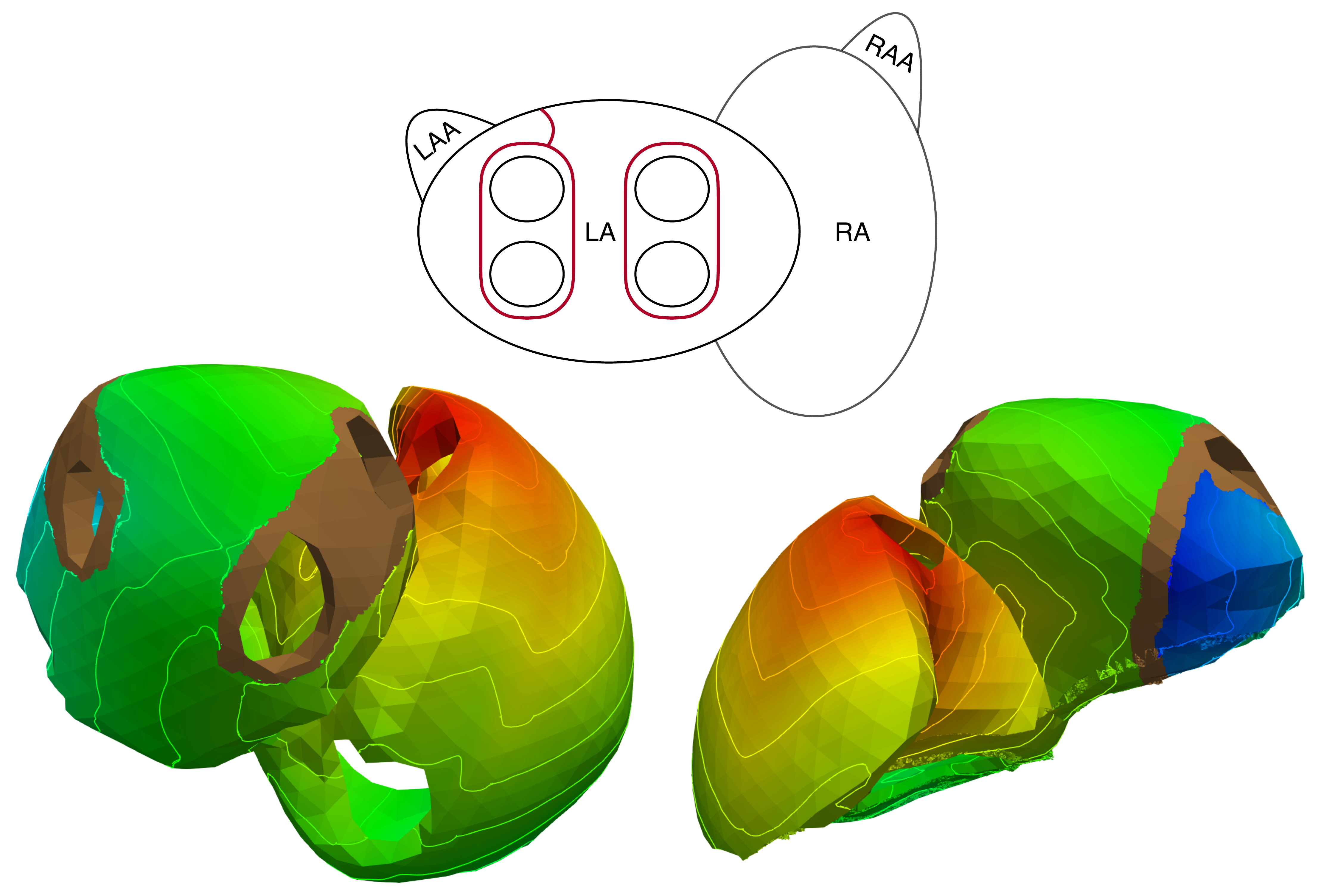}
        \caption{PVI+AL, 25\% red.}
    \end{subfigure}
    \begin{subfigure}[b]{0.32\textwidth}
        \includegraphics[width=\textwidth]{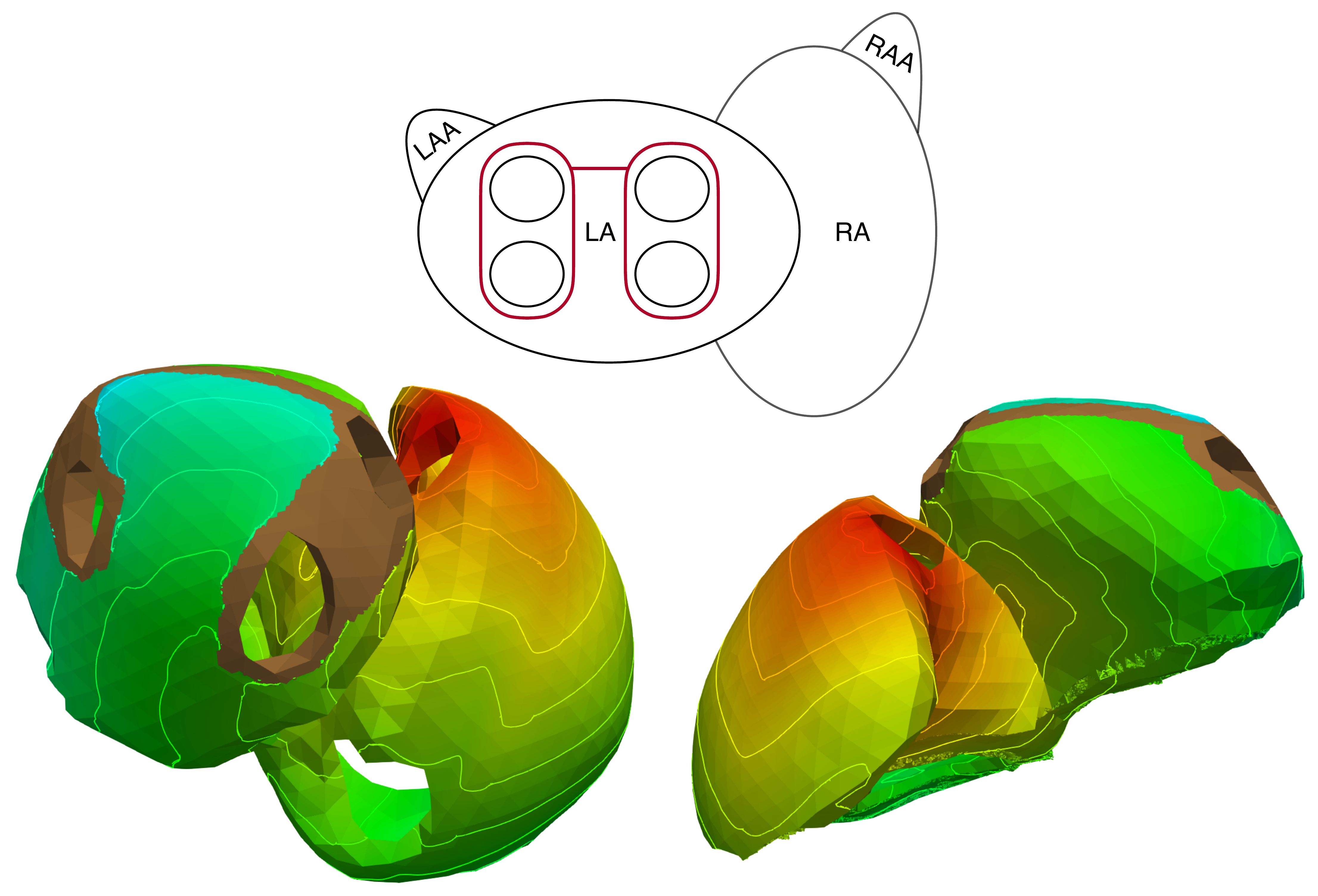}
        \caption{PVI+RL, 25\% red.}
    \end{subfigure}
    \begin{subfigure}[b]{0.32\textwidth}
        \includegraphics[width=\textwidth]{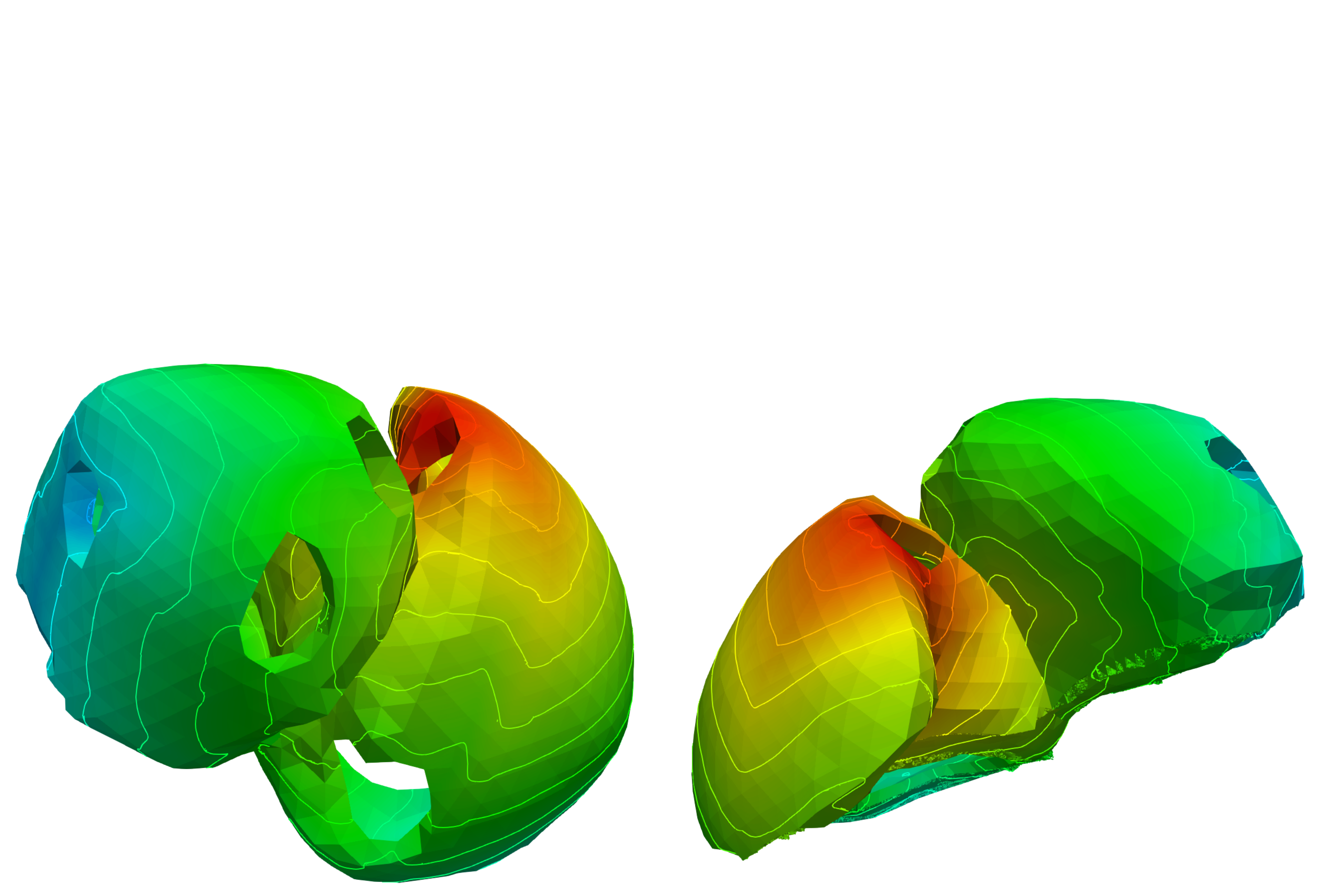}
        \caption{Control, 50\% red.}
    \end{subfigure}
    \begin{subfigure}[b]{0.32\textwidth}
        \includegraphics[width=\textwidth]{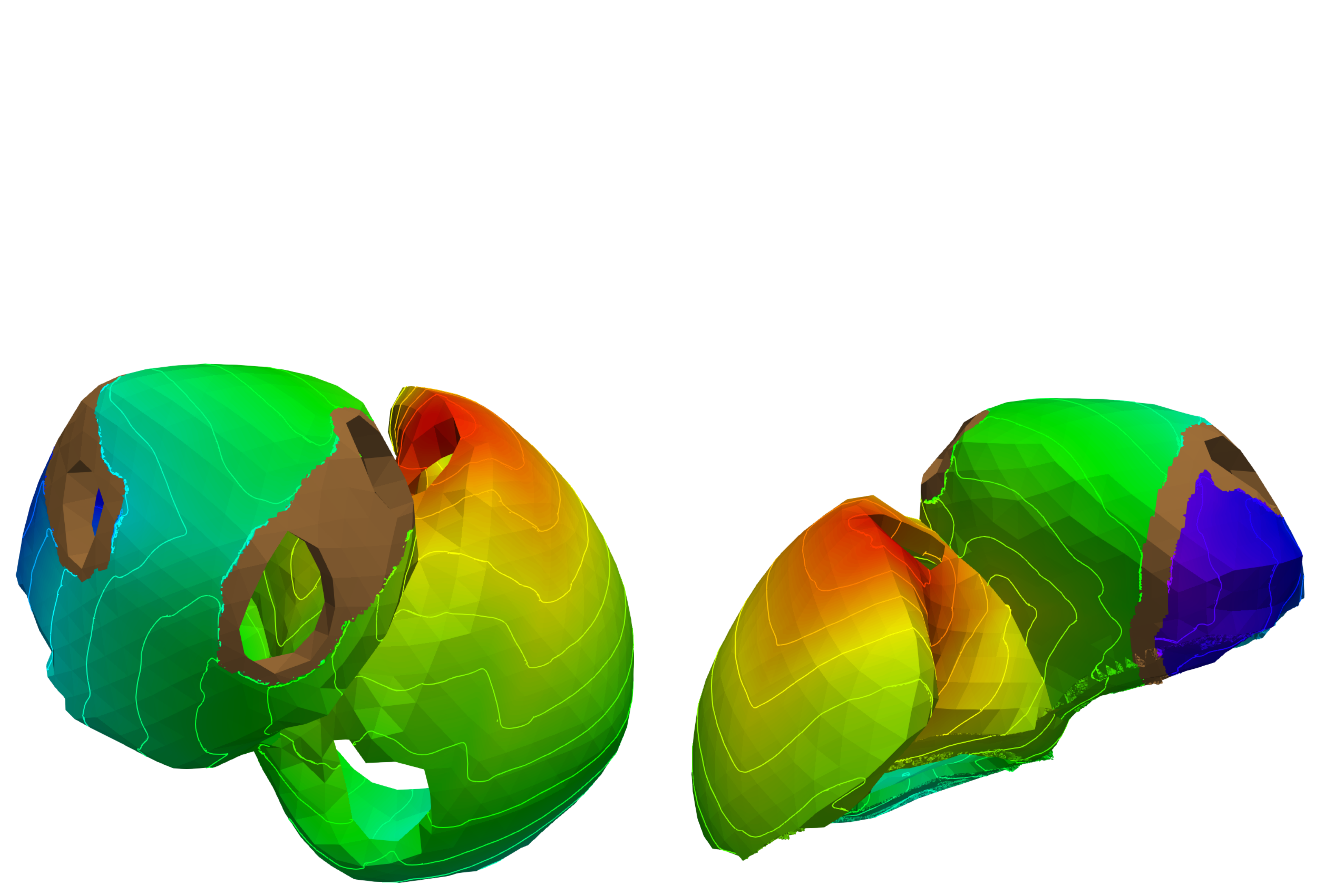}
        \caption{PVI+AL, 50\% red.}
    \end{subfigure}
    \begin{subfigure}[b]{0.32\textwidth}
        \includegraphics[width=\textwidth]{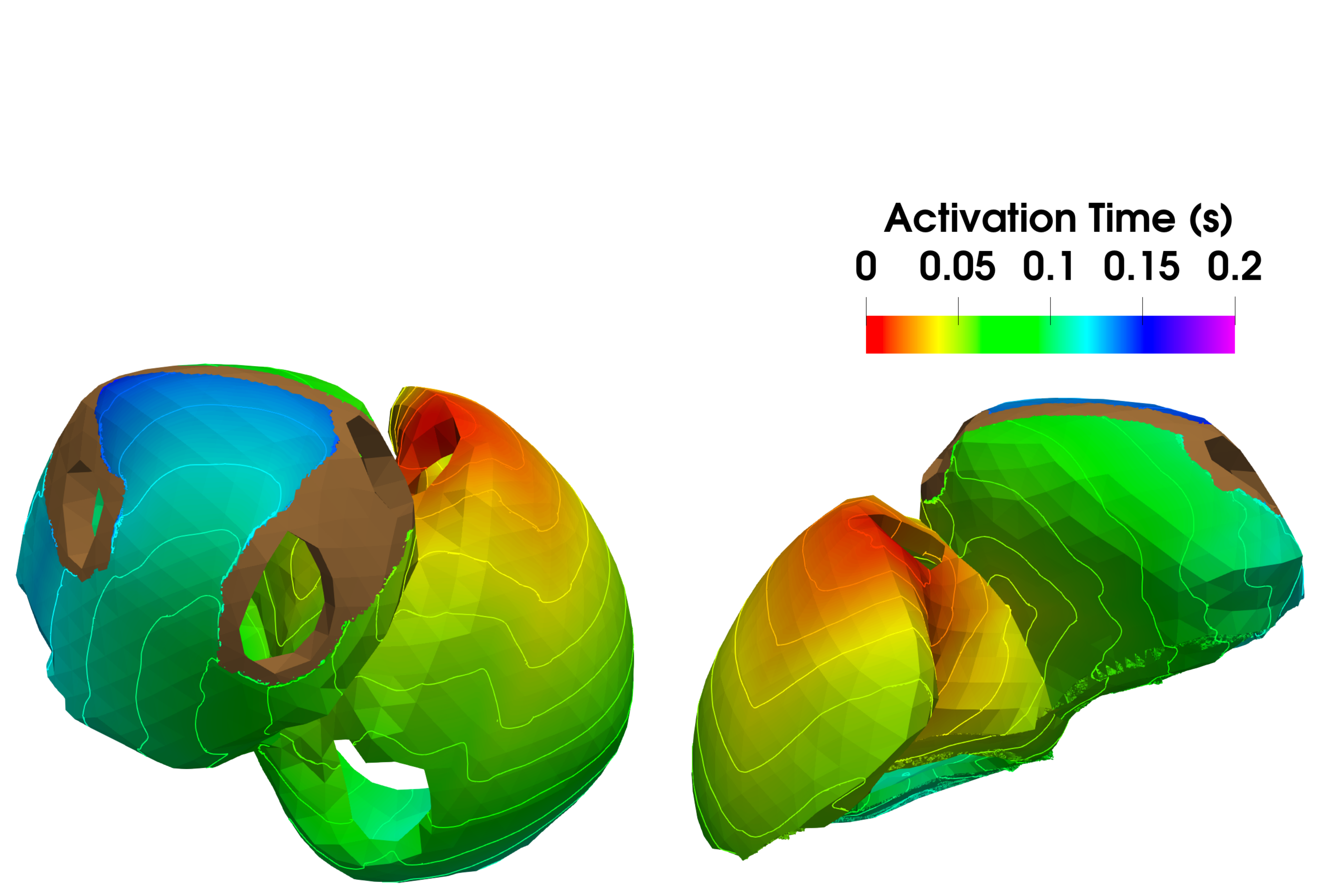}
        \caption{PVI+RL, 50\% red.}
    \end{subfigure}
    \caption{Activation time map and isochrones for the Control (a+d), the PVI+AL (b+e), and the PVI+RL (c+f) cases with a 25\% and 50\% reduction in conductivity.}
    \label{fig:reducedCV}
\end{figure*}

Atrial myocardium that was exposed to long lasting AF can appear to have a reduced conduction velocity due to the presence of fibrosis as a result of electrical and structural remodeling.
Therefore, six additional simulations were performed to investigate how a reduced conduction velocity in the atria affects the rest of the heart.
For the Control, PVI+AL, and PVI+RL cases the conductivity in both atria was reduced by 25\% and 50\%, respectively, to cover the diverse spectrum of atrial remodeling.
Figure~\ref{fig:reducedCV} shows the activation time maps for the three cases in both scenarios.
As expected, reduced conductivities result in a reduced conduction velocity and subsequently in a prolongation of atrial activation.
In the case of 50\% reduction in conductivity, this results in a 19\,ms delay of left atrial emptying in the three scenarios.
The change in ejected volume due to this delay is insignificant.
The delay in contraction does, however, affect blood pressure in the left atrium.
Especially in the case of an AL lesion, since the left atrial appendage takes up to 200\,ms to activate which in turn results in a temporal overlap of atrial and ventricular contraction.

%%%%%%%

\section{Discussion}
\label{discussion}

% short summary
We presented an electromechanically coupled four-chamber heart model including a closed-loop circulatory system that can be used to predict the acute hemodynamic effects of atrial ablation therapy.
We used patient specific measurements of AVPD and LV volume from MRI data to parameterize a healthy control case and performed a total of nine virtual ablations in the LA based on combinations of five standardized ablation lesions.
The impact of these scars on deformation and cardiovascular performance was evaluated using common biomarkers such as local activation time (LAT), EF, and AVPD.

% healthy control case
For the control case, we parameterized the active tension model for the atria and the ventricles such that the simulation results qualitatively matched the LV volume and AVPD evaluated from the volunteer's cine MRI data (see Fig.~\ref{fig:AVPD}).
Due to missing data from our healthy volunteer, the majority of parameters for the electrophysiological model and the circulatory system have been adapted from literature values of healthy subjects.
Hence, most clinical biomarkers can be reproduced faithfully.
With a root-mean-square error of $\mathrm{RMSE_{Vol}^{LV}} = 0.02$, $\mathrm{RMSE_{AVPD}^{LV}} = 0.38$\,mm, \linebreak $\mathrm{RMSE_{AVPD}^{RV}} = 0.99$\,mm for the normalized LV volume, mitral valve displacement, and tricuspid valve displacement, respectively, the simulated data match the experimental data well.
Therefore, we are confident that our model can reproduce physiological deformation patterns.

% atria systole
The ablation lesions in this work were modeled as perfectly isolating tissue.
Since there is a scarcity of data on elastomechanical properties of ablation scars in the atria, we increased the mechanical stiffness induced by an accumulation of collagen in the myocardium as it was observed in animal studies \citep{Jugdutt-1996-ID16489}.
Although these data are based on ventricular remodeling after myocardial infarction, we assume that it is a valid assumption, which is supported by the findings of \cite{takahashi07}.
They suggest that RFA scars behave elastomechanically like tissue in zones of chronic myocardial infarction.
Additionally, ablated tissue was modeled as isotropic with no preferred direction, since \cite{Caggiano-2021} showed that non-reinforced infarcts formed scars with no significant alignment of collagen fibers.
This should also be valid for ablations using electroporation or pulse field ablation, since the electrically isolating effect is the same as in RFA.
Mechanical changes of the tissue as a result of electroporation are less clear.
However, it most likely results in a stiffening of the tissue as well.
The increased stiffness resulted in a change of the left atrial EDPVR that did not only correlate with the amount of ablated tissue as one might expect (compare Fig.~\ref{fig:EDPVR} and Tab.~\ref{tab:LASystole}) but rather with a combination of the extent and position of the scars which is in agreement with ~\cite{Phung-2017}.

With the model presented in this manuscript being more comprehensive than the one published by \cite{Hormann-2017-ID12950} and \cite{Phung-2017}, our simulations confirm their results for the behavior during atrial systole.
In particular, our findings confirm a linear correlation between the change in left atrial EF and the amount of inactive tissue (see Fig.~\ref{fig:correlation}). 
This concurs with in vivo observations by \cite{takahashi07} in 40 patients who showed a reduced left atrial EF after RFA procedures ($18\pm11\%$).
Furthermore, a significant change in the activation sequence of the LA due to the ablation lesions AL and RL was observed.
Especially the delay in activation of the LAA through the AL lesion can be problematic since a large delay in LAA activation might lead to an overlap with LV contraction.
This was observed in our simulations in the form of a sudden increase in LA pressure at the beginning of LV contraction.
Such asynchronous contraction patterns are known to cause electrical and mechanical remodeling of the LA \citep{Sparks-1999-electrical,Sparks-1999-mechanical}.
The reduction in stroke volume of the LA results in a lower EDV of the LV.
However, this only accounts for maximally 1.8\% of LV stroke volume, which would likely not make a significant difference in well being for the patient.
The volunteer in this study was a young and healthy male and thus the atrial kick only contributed 11\% - 15\% to the EDV of the LV.
This contribution can be up to 40\% especially in older people \citep{Alhogbani13}.
Hence, a reduced atrial contraction would likely have a much higher impact on LV performance in these patients.

% ventricular systole
The increased stiffness of the scar tissue did have a measurable effect during ventricular systole.
We observed a reduced AVPD during both atrial and ventricular contraction.
The restricted movement of the valve plane resulted in reduced maximal volumes and increased maximal pressures of the LA during LV systole.
In the LV, the reduced stroke volume and valve plane movement result in a lower peak pressure during systole.
Overall, the right side of the heart is seemingly unaffected by the ablation scars in the LA.
Nevertheless, there are slight changes in volume during mid-diastole in the RV due to the adaptations of the closed-loop circulatory system to the changes in pre- and afterload.

\subsection{Limitations and perspectives}
\label{sec:limitations}

The focus of this study was to investigate the impact of ablation on the cardiovascular performance of the heart using a computational framework that accounts for the majority of known physiological mechanisms.
In order to do so, we had to make assumptions on some aspects of the model, which may affect the results:
\begin{itemize}
    \item[1.]
    Long-lasting AF can lead to electrical, contractile, and structural remodeling of atrial myocytes \citep{allessie02}.
    In this study, we assumed that all remodeling processes are reversed after a sufficient amount of time has passed.
    Therefore, the results presented here are more relevant for long-term performance changes due to ablation.
    \item[2.]
    We did not account for AF-related atrial fibrosis, which can lead to more complex propagation patterns.
    For this purpose, a biomechanical model of fibrosis needs to be developed first.
    \item[3.]
    Even though it is a valid approach to use linear tetrahedral elements to solve the monodomain equation \eqref{eq:PDEep}, it is known that this element type is prone to volumetric locking in solid mechanics problems.
    Nevertheless, a recent study has shown that errors in simulated hemodynamic outcomes are small enough when compared to observational uncertainties in clinical measurements \citep{Augustin-2021-ID16580}.
    In terms of motion, strains and stresses, \cite{Augustin-2021-ID16580} observed minor quantitative differences but essentially the same qualitative behavior.
    Hence, we decided to accept small errors due to volumetric locking in exchange for faster computation to make the amount of simulations required for this study feasible.
\end{itemize}

\section{Conclusion}
\label{conclusion}

This work provides an insight on how standard ablation strategies affect cardiovascular performance in a four-chamber heart model.
We confirmed the results by \cite{Hormann-2017-ID12950} and \cite{Phung-2017} for atrial systole on a different anatomical representation of the heart and extend the study based on preliminary work of \cite{busch13} by a physiologically valid closed-loop circulatory system model and an electromechanically coupled simulation framework for the whole heart.
With these additions, it is possible to show that ablation scars in the LA lower the amount of blood pumped into the LV depending on the extent of inactivated tissue.
Furthermore, we can observe an increase in pressure in the LA during ventricular contraction and a reduction in AVPD due to the increased stiffness in scar tissue.
These results support the hypothesis that extensive ablation not only negatively impacts cardiovascular performance but might also lead to further remodeling of the LA due to increased pressure and changes in the activation sequence.

\backmatter

\bmhead{Acknowledgments}
The authors would like to thank Olaf Dössel for valuable discussions.

\section*{Declarations}

\bmhead{Funding}
This research was funded by the Deutsche Forschungsgemeinschaft (DFG, German Research Foundation) – Project-ID 258734477 – SFB 1173.

\bmhead{Conflict of interest}
Tobias Gerach, Steffen Schuler, Andreas Wachter, and Axel Loewe declare that they have no conflict of interest.

\bmhead{Availability of data and materials}
The geometry and associated data used in this manuscript is publicly available \citep{geometry}.

\bmhead{Code availability}
Fiber generation algorithms and other useful tools can be found at https://github.com/KIT-IBT.

\begin{appendices}
\label{Appendix}

\section{Model parameters}

Here, we provide a complete list of model parameters that were used for all simulations in this study.
In particular, Table \ref{tab:EPParameter} contains the parameters for the electrophysiological model, Table \ref{tab:ActiveTension} refers to the parameters of the active stress model, and Tables \ref{tab:CircWholeHeartParameter} and \ref{tab:circInitial} contain the parameters for the circulatory system model.
Model parameters and initial conditions of the cellular electrophysiology model before pacing are adopted directly from the original publications~\citep{ohara11, courtemanche98}.

%%% EP parameters
\begin{table*}[t]
\caption{Electrophysiological parameters for the whole heart model.}
\label{tab:EPParameter}
\centering
\begin{tabular}{lrll}
\toprule
\textbf{Parameter} & \textbf{Value} & \textbf{Unit} & \textbf{Description} \\
\midrule
$(\sigma_\mathrm{f}, \sigma_\mathrm{s}, \sigma_\mathrm{n})$ & $(0.1979, 0.1046, 0.0363)$ & S/m & conductivities in ventricular bulk tissue \\
$(\sigma_\mathrm{f}, \sigma_\mathrm{s}, \sigma_\mathrm{n})$ & $(0.5756, 0.3042, 0.1046)$ & S/m & conductivities in ventricular fast conducting layer \\
$(\sigma_\mathrm{f}, \sigma_\mathrm{s}, \sigma_\mathrm{n})$ & $(1.0812, 0.1821, 0.1821)$ & S/m & conductivities in atrial bulk tissue \\
$(\sigma_\mathrm{f}, \sigma_\mathrm{s}, \sigma_\mathrm{n})$ & $(10^{-12},10^{-12},10^{-12})$ & S/m & conductivities in scar tissue \\
$\beta$ & $140000$ & 1/m & membrane surface-to-volume ratio \\
$C_\mathrm{m}$ & $0.01$ & F/m$^2$ & membrane capacitance \\
AV-delay & $0.160$ & s & atrio-ventricular conduction delay \\
Heart rate & $50$ & 1/60\,s & --\\
\bottomrule
\end{tabular}
\end{table*}

%%% Active stress parameters
\begin{table*}[t]
\caption{Active tension parameters for the whole heart model.}
\label{tab:ActiveTension}
\centering
\begin{tabular}{lrrll}
\toprule
\textbf{Parameter} & \multicolumn{2}{r}{\textbf{Value}} & \textbf{Unit} & \textbf{Description} \\
 & \textbf{Atria} & \textbf{Ventricle} & & \\
\midrule
$\lambda_0$ & 0.7 & 0.7 & - & minimum fiber stretch \\
$t_\text{emd}$ & 0.01 & 0.03 & s & electro-mechanical delay \\
$S_\text{peak}$ & 50 (RA), 80 (LA) & 900 (RV), 450 (LV) & kPa & peak isometric tension \\
$t_\text{dur}$ & 0.22 & 0.54 & s & duration of active contraction \\
$\tau_{c0}$ & 0.06 & 0.25 & s & base time constant of contraction \\
$\text{ld}$ & 5.0 & 5.0 & - & degree of length dependence \\
$\text{ld}_\text{up}$ & 0.5 & 0.5 & s & length dependence of upstroke time \\
$\tau_\text{r}$ & 0.06 & 0.08 & s & time constant of relaxation \\
$t_\text{cycle}$ & 1.2 & 1.2 & s & length of heart cycle \\
\bottomrule
\end{tabular}
\end{table*}

%%% Circ parameters
\begin{table*}[h!tb]
\caption{Circulatory system parameters for the whole heart model. Refer to \cite{Gerach-2021} for the system of equations.}
\label{tab:CircWholeHeartParameter}
\centering
\begin{tabular}{lrll}
\toprule
\textbf{Parameter} & \textbf{Value} & \textbf{Unit} & \textbf{Description} \\
\midrule
\multicolumn{4}{l}{\textit{Pulmonary and systemic circulation}} \\
$R_\text{SysArt}$ & 0.03 & $\text{mmHg}\cdot\text{s}\cdot\text{mL}^{-1}$ & systemic arterial resistance\\
$C_\text{SysArt}$ & 3.0 & $\text{mL}\cdot\text{mmHg}^{-1}$ & systemic arterial compliance \\
$V_\text{SysArtUnstr}$ & 800.0 & mL & unstressed systemic arterial volume \\
$R_\text{SysPer}$  & 0.6 & $\text{mmHg}\cdot\text{s}\cdot\text{mL}^{-1}$ & systemic peripheral resistance \\
$R_\text{SysVen}$  & 0.03 & $\text{mmHg}\cdot\text{s}\cdot\text{mL}^{-1}$ & systemic venous resistance \\
$C_\text{SysVen}$  & 150.0 & $\text{mL}\cdot\text{mmHg}^{-1}$ & systemic venous compliance \\
$V_\text{SysVenUnstr}$  & 2850.0 & mL & unstressed systemic venous resistance \\
$R_\text{PulArt}$  & 0.02 & $\text{mmHg}\cdot\text{s}\cdot\text{mL}^{-1}$ & pulmonary arterial resistance \\
$C_\text{PulArt}$  & 10.0 & $\text{mL}\cdot\text{mmHg}^{-1}$ & pulmonary arterial compliance \\
$V_\text{PulArtUnstr}$  & 150.0 & mL & unstressed pulmonary arterial volume  \\
$R_\text{PulPer}$  & 0.07 & $\text{mmHg}\cdot\text{s}\cdot\text{mL}^{-1}$ & pulmonary peripheral resistance \\
$R_\text{PulVen}$  & 0.03 & $\text{mmHg}\cdot\text{s}\cdot\text{mL}^{-1}$ & pulmonary venous resistance \\
$C_\text{PulVen}$  & 15.0 & $\text{mL}\cdot\text{mmHg}^{-1}$ & pulmonary venous compliance \\
$V_\text{PulVenUnstr}$ & 200.0 & mL & unstressed pulmonary venous volume \\
$\rho_\text{Blood}$ & 1060 & kg/m$^3$ & blood density \\
\midrule
\multicolumn{4}{l}{\textit{Mitral valve}} \\
$A_\text{Ref}$ & 15.0 & $\text{cm}^2$ & reference area \\
$M_\text{max}$ & 0.7 & - & maximum area ratio \\
$M_\text{min}$ & 0.001 & - & minimum area ratio \\
$K_\text{o}$ & 15.0 & $\text{mmHg}^{-1}\cdot\text{s}^{-1}$ & opening rate coefficient \\
$K_\text{c}$ & 3.0 & $\text{mmHg}^{-1}\cdot\text{s}^{-1}$ & closing rate coefficient \\
\midrule
\multicolumn{4}{l}{\textit{Tricuspid valve}} \\
$A_\text{Ref}$ & 15.0 & $\text{cm}^2$ & reference area \\
$M_\text{max}$ & 0.7 & - & maximum area ratio \\
$M_\text{min}$ & 0.001 & - & minimum area ratio \\
$K_\text{o}$ & 20.0 & $\text{mmHg}^{-1}\cdot\text{s}^{-1}$ & opening rate coefficient \\
$K_\text{c}$ & 6.0 & $\text{mmHg}^{-1}\cdot\text{s}^{-1}$ & closing rate coefficient \\
\midrule
\multicolumn{4}{l}{\textit{Aortic valve}} \\
$A_\text{Ref}$ & 7.0 & $\text{cm}^2$ & reference area \\
$M_\text{max}$ & 0.95 & - & maximum area ratio \\
$M_\text{min}$ & 0.001 & - & minimum area ratio \\
$K_\text{o}$ & 10.0 & $\text{mmHg}^{-1}\cdot\text{s}^{-1}$ & opening rate coefficient \\
$K_\text{c}$ & 6.0 & $\text{mmHg}^{-1}\cdot\text{s}^{-1}$ & closing rate coefficient \\
\midrule
\multicolumn{4}{l}{\textit{Pulmonary valve}} \\
$A_\text{Ref}$ & 7.0 & $\text{cm}^2$ & reference area \\
$M_\text{max}$ & 0.7 & - & maximum area ratio \\
$M_\text{min}$ & 0.001 & - & minimum area ratio \\
$K_\text{o}$ & 20.0 & $\text{mmHg}^{-1}\cdot\text{s}^{-1}$ & opening rate coefficient \\
$K_\text{c}$ & 10.0 & $\text{mmHg}^{-1}\cdot\text{s}^{-1}$ & closing rate coefficient \\
\bottomrule
\end{tabular}
\end{table*}

\begin{table*}[htb]
    \caption{Initial values for the circulatory system model.}
    \label{tab:circInitial}
    \centering
    \begin{tabular}{lrll}
    \toprule
    \textbf{Parameter} & \textbf{Value} & \textbf{Unit} & \textbf{Description} \\
    \midrule
    \multicolumn{4}{l}{\textit{Initial conditions}} \\
    $V_\text{Tot}$ & 5680.0 & mL & total blood volume \\
    $V_\text{SysArt}$ & 1064.2 & mL & systemic arterial volume \\
    $V_\text{PulArt}$ & 294.16 & mL & pulmonary arterial volume \\
    $V_\text{PulVen}$ & 344.39 & mL & pulmonary venous volume \\
    $p_\text{LV}$ & 8.0 & mmHg & left ventricular pressure \\
    $p_\text{LA}$ & 8.0 & mmHg & left atrial pressure \\
    $p_\text{RV}$ & 4.0 & mmHg & right ventricular pressure \\
    $p_\text{RA}$ & 4.0 & mmHg & right atrial pressure \\
    $Q_\text{SysArt}$ & 0.0 & mL$\cdot \text{s}^{-1}$ & aortic valve flow \\
    $Q_\text{PulArt}$ & 0.0 & mL$\cdot \text{s}^{-1}$ & pulmonary valve flow \\
    $Q_\text{Mitral}$ & 0.0 & mL$\cdot \text{s}^{-1}$ & mitral valve flow \\
    $Q_\text{Tricuspid}$ & 0.0 & mL$\cdot \text{s}^{-1}$ & tricuspid valve flow \\
    $\sigma_\text{SysArt}$ & 0.0 & - & aortic valve state \\
    $\sigma_\text{PulArt}$ & 0.0 & - & pulmonary valve state \\
    $\sigma_\text{Mitral}$ & 0.0 & - & mitral valve state \\
    $\sigma_\text{Tricuspid}$ & 0.0 & - & tricuspid valve state \\
    \bottomrule
    \end{tabular}
\end{table*}

\end{appendices}


\bibliography{sn-article}% common bib file


\end{document}